\documentclass[12pt]{article}
\usepackage{putex}

\usepackage[T1]{fontenc} 				% Use 8-bit encoding that has 256 glyphs

\usepackage{amssymb,amsmath,amscd}
\usepackage[english]{babel}
\usepackage{titlesec}
\usepackage{slashed}
\usepackage{hyperref}
\usepackage{graphicx,xcolor}
\usepackage{enumitem}
\usepackage{multirow}
\usepackage[absolute]{textpos}
\usepackage[margin=1.5em,labelfont=bf]{caption}
\usepackage{cite}

\hypersetup{
	colorlinks=true,
	linkcolor=dark-blue,
	citecolor=dark-red,
	urlcolor=dark-green,
	linktoc=all
}

\graphicspath{{./figures/}}

\usepackage{geometry} 					% Document margins
\numberwithin{equation}{section} 

\titleformat{\section}[block]{\Large\bfseries}{\thesection}{1em}{} 
\titleformat{\subsection}[block]{\bfseries}{\thesubsection}{1em}{} 
\titlespacing*{\section}{0pt}{1em}{1em}
\titlespacing*{\subsection}{0pt}{0.75em}{0.75em}

\graphicspath{{./figures/}}

\newcommand\Tstrut{\rule{0pt}{2.6ex}}         % = `top' strut
\definecolor{dark-gray}{gray}{0.20}
\definecolor{gray}{gray}{0.30}
\definecolor{light-gray}{gray}{0.80}
\definecolor{dark-red}{rgb}{0.7,0,0}
\definecolor{dark-green}{rgb}{0.1,0.4,0}
\definecolor{dark-blue}{rgb}{0.3,0.3,0.7}
\definecolor{light-blue}{rgb}{0.8,0.8,1}
\definecolor{cardinal}{rgb}{0.6,0,0}
\definecolor{darkgreen}{rgb}{0,0.5,0}
\definecolor{golden}{rgb}{0.92, 0.7, 0}
\definecolor{midnight}{rgb}{0, 0, 0.5}
\definecolor{darkblue}{rgb}{0.2, 0, 0.8}
\definecolor{forestgreen}{rgb}{0.13, 0.55, 0.13}

\newcommand\bK{\mathbf{K}}
\newcommand\bD{\mathbf{D}}
\newcommand\bM{\mathbf{M}}
\newcommand\bH{\mathbf{H}}
\newcommand\bC{\mathbf{C}}
\newcommand\bP{\mathbf{P}}
\newcommand\bR{\mathbf{R}}

\newcommand\bQ{\mathbf{Q}}
\newcommand\cA{\mathcal{A}}

\newcommand\cC{\mathcal{C}}
\newcommand\cD{\mathcal{D}}

\newcommand\cF{\mathcal{F}}
\newcommand\cG{\mathcal{G}}
\newcommand\cH{\mathcal{H}}

\newcommand\cJ{\mathcal{J}}

\newcommand\cL{\mathcal{L}}

\newcommand\cN{\mathcal{N}}
\newcommand\cO{\mathcal{O}}

\newcommand\cQ{\mathcal{Q}}
\newcommand\cR{\mathcal{R}}
\newcommand\cS{\mathcal{S}}
\newcommand\cT{\mathcal{T}}

\newcommand\cV{\mathcal{V}}
\newcommand\cW{\mathcal{W}}

\newcommand\cZ{\mathcal{Z}}

\newcommand\fg{\mathfrak{g}}
\newcommand\fh{\mathfrak{h}}

\newcommand{\dd}{\mathrm{d}}
\newcommand{\e}{\mathrm{e}}

\newcommand\Tr{\mathrm{Tr}\,}

\newcommand{\f}[2]{\frac{#1}{#2}}
\newcommand{\comm}[2]{\left[{#1},{#2}\right]}
\newcommand{\acomm}[2]{\left\{{#1},{#2}\right\}}

\newcommand{\nn}{\nonumber}
\newcommand{\ds}{{\rm d}s}
\newcommand{\wti}[1]{\widetilde{#1}}

\newcommand {\be} {\begin {equation}}
\newcommand {\ee} {\end {equation}}
\newcommand {\bes} {\begin {equation*}}
\newcommand {\ees} {\end {equation*}}

\newcommand{\es}[2] {\begin{equation} \label{#1} \begin{split} #2 \end{split} \end{equation}}

\newcommand\SO{\mathrm{SO}}

\newcommand\UU{\mathrm{U}}
\newcommand\SU{\mathrm{SU}}

\newcommand\USp{\mathrm{USp}}
\newcommand\OSp{\mathrm{OSp}}

\newcommand\su{\mathfrak{su}}
\newcommand\psu{\mathfrak{psu}}

\newcommand\osp{\mathfrak{osp}}

\newcommand\uu{\mathfrak{u}}

\begin{document}

\preprint{PUPT-2629}

\institution{PU}{Joseph Henry Laboratories, Princeton University, Princeton, NJ 08544, USA}
\institution{Padova}{Universit\`a di Padova, Dipartimento di Fisica e Astronomia,\cr via Marzolo 8, 35131 Padova, Italy}
\institution{PadovaAgain}{INFN, Sezione di Padova, via Marzolo 8, 35131 Padova, Italy}

\title{
\vspace{-0.5in}
\textbf{One-Dimensional Sectors From the\\ Squashed Three-Sphere}
}

\authors{Pieter Bomans\worksat{\PU,\Padova,\PadovaAgain} and Silviu S.~Pufu\worksat{\PU}}

\abstract{
\noindent 
Three-dimensional ${\cal N} = 4$  superconformal field theories contain 1d topological sectors consisting of twisted linear combinations of half-BPS local operators that can be inserted anywhere along a line.  After a conformal mapping to a round three-sphere, the 1d sectors are now defined on a great circle of $S^3$.  We show that the 1d topological sectors are preserved under the squashing of the sphere.  For gauge theories with matter hypermultiplets,  we use supersymmetric localization to derive an explicit description of the topological sector associated with the Higgs branch.  Furthermore, we find that the dependence of the 1d correlation functions on the squashing parameter $b$ can be removed after appropriate rescalings.  One can introduce real mass and Fayet-Iliopolous parameters that, after appropriate rescalings, modify the 1d theory on the squashed sphere precisely as they do on the round sphere.    In addition, we also show that when a generic 3d $\cN=4$ theory is deformed by real mass parameters, this deformation translates into a universal deformation of the corresponding 1d theory.
}
\date{December 2021}

\maketitle

{
	\hypersetup{linkcolor=black}
	\tableofcontents
}

\section{Introduction}
\label{sec:introduction}

Following the work of Pestun \cite{Pestun:2007rz}, there has been a plethora of exact results obtained using the technique of supersymmetric localization  \cite{Witten:1988ze}, in various dimensions and with various amounts of supersymmetry (see \cite{Pestun:2016zxk} for a collection of reviews and references). These exact results are mostly for partition functions of supersymmetric field theories on curved manifolds and/or for expectation values of supersymmetry-preserving non-local operators, such as Wilson loops, 't Hooft loops, integrated local operators, etc\@.   Such observables generally depend on only a few of the parameters that define the theory.  For instance, in three dimensions, the smallest amount of supersymmetry where such exact computations are possible is ${\cal N} = 2$, corresponding to four real supercharges in flat space. The dependence of the partition function on the parameters of the supersymmetric background was studied in \cite{Closset:2012ru,Closset:2013vra}.  Particularly relevant to the present work are gauge theories coupled to matter placed on round or squashed three-spheres \cite{Kapustin:2009kz,Jafferis:2010un,Hama:2010av,Hama:2011ea,Imamura:2011wg}, where the partition function depends on the squashing, the real masses, and the Fayet-Iliopolous (FI) parameters.  

Since the observables mentioned above only depend on a limited set of parameters, they capture only a limited amount of information.  On the other hand, quantum field theories possess much richer classes of observables. Of particular interest are the local operators, which are the central object of study in the conformal bootstrap program (for reviews, see \cite{Rychkov:2016iqz,Simmons-Duffin:2016gjk,Poland:2018epd,Chester:2019wfx,Qualls:2015qjb}). In general, correlation functions of local operators at separated points are not supersymmetric, and therefore it is not possible to directly calculate them using supersymmetric localization.  In some cases, however, correlation functions of certain local operators can be determined using supersymmetric localization inputs, provided that the symmetries of the theory are restrictive enough to fix the position-dependence of the correlation functions of interest up to a few undetermined parameters. This is the case, for example, for the two-point functions of conserved currents or of the stress-energy tensor in 3d ${\cal N} = 2$ superconformal field theories (SCFTs) \cite{Closset:2012ru,Closset:2012vg}.\footnote{See also \cite{Gerchkovitz:2016gxx} for computations of Coulomb branch operators in ${\cal N} = 2$ SCFTs in 4d.}  Another example is the four-point function of stress tensor multiplet operators in holographic theories in three and four dimensions \cite{Binder:2018yvd,Binder:2019mpb,Binder:2019jwn,Binder:2020ckj,Chester:2020dja}.

The goal of this paper is to study a new case in which one can calculate certain correlation functions of local operators directly using supersymmetric localization.  We study  correlators of Higgs and Coulomb branch operators of ${\cal N} = 4$ supersymmetric theories on a squashed three-sphere. In \cite{Beem:2013sza,Chester:2014mea,Beem:2016cbd,Dedushenko:2016jxl}, it was noticed that all 3d ${\cal N}=4$ superconformal field theories contain topological 1d sectors comprised of ``twisted'' $1/2$-BPS operators (i.e.~$1/2$-BPS operators whose R-symmetry indices are contracted with certain space-dependent polarization vectors).\footnote{These 1d topological sectors are 3d analogs of the chiral algebra sector of 4d ${\cal N}= 2$  and 6d $(2, 0)$ SCFTs introduced in \cite{Beem:2013sza} and \cite{Beem:2014kka} , respectively.}  In particular, when such operators are inserted on a line, their correlation functions are topological in the sense that they depend only on the ordering of the operators on the line and not on the separation between the insertions. These 1d topological sectors provide a deformation quantization of the Higgs or Coulomb branch of these theories, as was further studied in \cite{Beem:2016cbd,Etingof:2019guc,Etingof:2020fls}.\footnote{See \cite{Dedushenko:2019mzv,Dedushenko:2019mnd} for the relation of deformation quantization to the VOAs associated to 4d $\cN=2$ SCFTs.}  In \cite{Dedushenko:2016jxl,Dedushenko:2017avn,Dedushenko:2018icp}, it was further shown that in 3d ${\cal N}=4$ SCFTs constructed as infrared limits of gauge theories coupled to hypermultiplet matter, the 1d sectors can be accessed using supersymmetric localization. For instance, in the Higgs branch case studied in \cite{Dedushenko:2016jxl}, it was shown that, after a conformal map to $S^3$, the 3d theory localizes to a 1d theory on a great circle of $S^3$. This 1d theory can be written as a topological gauged quantum mechanics, with the matter fields being anti-periodic scalars on the circle.  After gauge fixing, the topological gauged quantum mechanics can be equivalently recast as a 1d Gaussian theory coupled to a matrix model. This 1d theory can be further modified by introducing real mass and/or FI parameters for the 3d theory on $S^3$.  While the FI parameters retain the topological nature of the 1d theory, in the presence of the real mass parameters, some of the correlation functions acquire a (relatively simple) position dependence. 

In this paper, we take these constructions one step further and show that the 1d sectors mentioned above persist on a squashed three-sphere.\footnote{In \cite{Panerai:2020boq}, a similar extension for theories with hypermultiplets coupled to background vector multiplets was discussed from a slightly different point of view, using equivariant cohomology. To illustrate their analysis, Ref.~\cite{Panerai:2020boq} discussed a 1d sector of $\cN=4$ theories on $S^1\times S^2$.}  For the more general case of ${\cal N} = 2$ theories, it is known that there are several kinds of squashed sphere backgrounds, distinguished by the different choices of couplings that are required to preserve supersymmetry.  Among the more symmetric backgrounds, there exists an $\mathfrak{su}(2) \times \mathfrak{u}(1)$-invariant squashing for which the partition function is independent of the squashing parameter $b$, as well as $\mathfrak{su}(2) \times \mathfrak{u}(1)$-invariant and $\mathfrak{u}(1) \times \mathfrak{u}(1)$-invariant squashings for which the partition function does depend non-trivially on $b$ \cite{Hama:2010av,Hama:2011ea,Imamura:2011wg}.  A supersymmetric theory with ${\cal N}=2$ supersymmetry would then preserve the superalgebra $\mathfrak{su}(2|1) \times \mathfrak{u}(1)$, which contains four real supercharges.  For an ${\cal N} = 4$ theory, the  $\mathfrak{su}(2|1) \times \mathfrak{u}(1)$ algebra is extended to either $\mathfrak{su}(2|1) \times \mathfrak{psu}(1|1)$, and therefore it preserves six real supercharges, or to $\mathfrak{psu}(2|2) \times \mathfrak{u}(1)$, and therefore it preserves eight real supercharges \cite{Minahan:2021pfv}.

A standard way of constructing supersymmetric field theories in curved space is to couple a flat space theory to off-shell conformal or Poincar\'e supergravity, and give the supergravity fields expectation values that preserve supersymmetry \cite{Festuccia:2011ws,Hama:2012bg,Dumitrescu:2012ha}. For ${\cal N} = 2$ theories on a squashed three-sphere, the values of the supergravity fields that correspond to a supersymmetric squashed three-sphere were identified in \cite{Closset:2012ru}.   In the recent work \cite{Minahan:2021pfv}, the authors started with ${\cal N} = 2$ supergravity in 4d and constructed supersymmetric theories on a squashed three-sphere by performing a circle reduction.  They identified a family of backgrounds that generically preserve ${\cal N} = 2$ supersymmetry, but for special values of the parameters the symmetry is enhanced to either $\mathfrak{su}(2|1) \times \mathfrak{psu}(1|1)$ or $\mathfrak{psu}(2|2) \times \mathfrak{u}(1)$, thus corresponding to 3d ${\cal N} = 4$ theories on a squashed three-sphere.  The downside of this approach, however, is that one cannot easily construct 3d theories with twisted vector multiplets and hypermultiplets.

Inspired by the work of \cite{Minahan:2021pfv}, we construct ${\cal N} = 4$ theories on a squashed three-sphere by starting with ${\cal N} = 4$ off-shell conformal supergravity \cite{Banerjee:2015uee,Butter:2013rba,Butter:2013goa} coupled to matter and giving appropriate expectation values to the fields in the supergravity multiplet. As also found in \cite{Minahan:2021pfv}, there are two different backgrounds that preserve $\mathfrak{su}(2|1) \times \mathfrak{psu}(1|1)$ which are related to each other through mirror symmetry. In particular, the ${\cal N} = 4$ conformal supergravity has $\mathfrak{su}(2)_H \times \mathfrak{su}(2)_C$ R-symmetry, but our backgrounds preserve only a $\mathfrak{u}(1)_H \times \mathfrak{u}(1)_C$ Cartan subalgebra.  The two $\mathfrak{su}(2|1) \times \mathfrak{psu}(1|1)$ backgrounds can be related by mirror symmetry, which interchanges the roles of $\mathfrak{u}(1)_H$ and $\mathfrak{u}(1)_C$.  In addition to the $\mathfrak{su}(2|1) \times \mathfrak{psu}(1|1)$-preserving backgrounds, we also find two $\mathfrak{psu}(2|2) \times \mathfrak{u}(1)$-preserving background, similarly related by mirror symmetry, as expected from \cite{Minahan:2021pfv}.

For a given $\mathfrak{su}(2|1) \times \mathfrak{psu}(1|1)$-preserving background, we show that there exist two 1d sectors (the two sectors get trivially interchanged if we consider the other background) where operators are inserted along a geodesic circle of the squashed sphere. In the following we will denote these 1d theories by $\cT_H$ and $\cT_C$, as they are related to the $\uu(1)_H$ and $\uu(1)_C$ R-symmetry, respectively. For gauge theories built out of vector multiplets and hypermultiplets, $\cT_H$ contains ``Higgs branch operators'' while $\cT_C$ contains ``Coulomb branch operators,'' just as on the round sphere \cite{Dedushenko:2016jxl,Dedushenko:2017avn,Dedushenko:2018icp}.  Following an argument similar to the one in \cite{Dedushenko:2016jxl}, we further show that the 1d sector $\cT_H$ is described by a rather trivial modification of the gauged quantum mechanics from the round sphere case.

As a preview, for a 3d gauge theory with gauge group $G$ and a hypermultiplet transforming in representation $\cR_H$ of $G$, the partition function of the 3d theory can be written as
\begin{equation}\label{ZbIntro}
	Z_{S_b^3}(m) = \f{1}{|\cW|}\int_{\fh} \dd \sigma\,  \Delta_b(\sigma)Z_\text{1d} (\sigma) \,,
\end{equation}
where $|\cW|$ is the order of the Weyl group of the gauge algebra $\mathfrak{g} = \text{Lie}(G)$, and $\fh$ is the Cartan subalgebra of $\mathfrak{g}$. The one-loop determinant $\Delta_b$ is given by 
\begin{equation}\label{DeltabDef}
	\Delta_b(\sigma) = \frac{1}{b^{|G|}}{\det}_\text{adj}^\prime\, 2 \sinh \f{\pi\sigma}{b} \,,
\end{equation}
with the determinant being taken in the adjoint representation\footnote{The prime indicates that the Cartan elements are excluded.} and $|G|$ being the rank of the gauge group. Finally, the partition function of the 1d Gaussian theory is given by
\begin{equation}\label{Gaussian1d}
	Z_\text{1d}(\sigma) = \int \prod_I \cD \cZ^I\,\,  \exp\left\{ 4\pi r \, \int \dd\alpha  \left( b \cZ_I\partial_\alpha \cZ^I + \cZ_I \left(\sigma_a T^a \right)^I{}_J \cZ^J \right)\right\}\,.
\end{equation}
In this formula, $\alpha \in [0, 2 \pi)$ is the one-dimensional coordinate parameterizing a circle on the squashed sphere, $\cZ^I$ and $\cZ_I$ are anti-periodic bosonic fields\footnote{The indices are raised and lowered with an anti-symmetric tensor $\varepsilon_{IJ}$.} related to the hypermultiplet scalars, $T^a$ are the generators of the Lie algebra $\fg$ which act in the appropriate representation $\cR_H$.  One can use this 1d theory to calculate correlation functions of gauge-invariant products of $\cZ^I$.  Rescaling $\cZ^I \to \frac{1}{\sqrt{b}} \cZ^I$ and $\sigma \to b \sigma$ yields a 1d theory independent of $b$.  The above results can straightforwardly be extended to include FI and real mass parameters, as will be described in more detail in the main text.  Performing the Gaussian integral over the hypermultiplet fields, one obtains the matrix model for a squashed sphere, whose independence of $b$ was noticed in \cite{Chester:2021gdw,Minahan:2021pfv}. 

The remainder of the paper is organized as follows. In Section \ref{sec:background}, we start by introducing the $\cN=4$ squashed sphere backgrounds, study the supersymmetry algebra they preserve, and formulate the QFTs we will study in these backgrounds. Having defined the QFTs of interest we proceed to show how one can construct two one-dimensional sectors within such QFTs. In Section \ref{sec:1dSectorCohomology}, we present a cohomological construction of these sectors. In Section \ref{sec:1dSector}, we use supersymmetric localization to derive an explicit description of the 1d $\cT_H$ sector of 3d theories constructed from vector multiplets coupled to hypermultiplets.   Next, in Section \ref{sec:genericMassDef}, we again take a more general perspective to show that for any  $\cN=4$ QFT on the squashed sphere real mass deformations correspond to deformations of the 1d $\cT_H$ sector. We end with a discussion of our results in Section~\ref{sec:discussion}.   Conventions and several technical details are relegated to the Appendices.

%%%%%%%%%%%%%%%%%%%%%%%%%%%%%
\section{$\cN=4$ theories on the squashed sphere}
\label{sec:background}
%%%%%%%%%%%%%%%%%%%%%%%%%%%%%

Let us start by explaining how to formulate $\cN=4$ supersymmetric quantum field theories on a squashed three-sphere. As mentioned in the Introduction, a general procedure to construct supersymmetric QFTs on curved manifolds is to consider the action and supersymmetry variations of the matter multiplets coupled to off-shell supergravity, and subsequently freeze the fields in the Weyl multiplet to supersymmetric configurations  \cite{Festuccia:2011ws,Hama:2012bg,Dumitrescu:2012ha}. The choice of off-shell supergravity theory determines whether a background with the desired properties exists and at the same time limits the possible terms in the matter Lagrangians one can write down.

In this work we are interested in considering QFTs on the squashed sphere, where the metric preserves an $\mathfrak{su}(2) \times \mathfrak{u}(1)$ isometry. In particular, we choose the metric of the squashed sphere to be
\begin{equation}\label{3dmetric}
	\ds_3^2 = \f{r^2}{4}\left( \dd\theta^2 + \sin^2\theta\dd\phi^2 + \left(\f{b+b^{-1}}{2}\right)^2\left( \dd\psi+\cos\theta\dd\phi \right)^2 \right)\,.
\end{equation}
In this expression, $b$ is the squashing parameter with $b=1$ corresponding to the round sphere, $r$ is the radius of the sphere, and the angles $\phi$ and $\psi$ obey the periodic identifications $\phi\sim \phi + 2\pi$, $\psi\sim \psi + 4\pi$, while the range of $\theta$ is $[0, \pi]$.  This metric, combined with appropriate choices for the other supergravity background fields, can be coupled to a variety of matter multiplets. In this work we will focus on theories built out of hypermultiplets and vector multiplets, as well as their twisted analogs. The details about the structure of such multiplets will be reviewed below. 

In the remainder of this section we will introduce several ways to couple our background to such matter fields where the possible matter coupling are constrained by the choice of off-shell supergravity theory. In the following list we summarize the various possibilities as well as their limitations.
\begin{itemize}
	\item \textit{Conformal supergravity} (consisting of a Weyl multiplet) \cite{Banerjee:2015uee,Butter:2013rba,Butter:2013goa}: We will introduce supersymmetric backgrounds for the Weyl multiplet for which the metric takes the form \eqref{3dmetric}. These backgrounds preserve either an $\mathfrak{su}(2|1) \times \mathfrak{psu}(1|1)$ or $\mathfrak{psu}(2|2) \times \mathfrak{u}(1)$ superalgebra. In conformal supergravity one is limited to considering only conformal matter on these backgrounds. In particular, one can consider kinetic terms for hypermultiplets and twisted hypermultiplets as well as mixed abelian Chern-Simons terms. However, one cannot consider non-conformal terms such as Yang-Mills (YM) terms for dynamical vector multiplets, and for this reason such a background will not be sufficient for our purposes.
	\item \textit{Conformal supergravity with a compensating vector multiplet}:  In this theory, we will still be able to find the $\mathfrak{su}(2|1) \times \mathfrak{psu}(1|1)$ and $\mathfrak{psu}(2|2) \times \mathfrak{u}(1)$-preserving backgrounds mentioned above, even after giving non-zero values to the fields in the compensating multiplet. In addition to the interactions allowed in the previous bullet point, we will now also be able to add Yang-Mills terms for the vector multiplets, but not for the twisted vector multiplets.
	\item \textit{Conformal supergravity with a compensating twisted vector multiplet}: This case is mirror dual to the previous one and hence we will again be able to find the $\mathfrak{su}(2|1) \times \mathfrak{psu}(1|1)$ and $\mathfrak{psu}(2|2) \times \mathfrak{u}(1)$-preserving backgrounds mentioned above, even after giving non-zero values to the fields in the compensating multiplet. In addition to the interactions allowed in the first bullet point, we will now be able to add Yang-Mills terms for the twisted vector multiplets, but not for the vector multiplets.
	\item \textit{Conformal supergravity with both a compensating vector and twisted vector multiplet}: This theory only contains the $\mathfrak{su}(2|1) \times \mathfrak{psu}(1|1)$-invariant squashed sphere as a background. In addition to the interactions allowed in the first bullet point, we will now be able to add Yang-Mills terms for both vector multiplets and twisted vector multiplets.
	\item \textit{Background vector and twisted vector multiplets}: For any of the theories in this list we can add additional background vector or twisted vector multiplets that couple to global symmetries of the matter theory. By giving supersymmetry-preserving expectation values to the fields in these vector multiplets, one can introduce real mass parameters and FI parameters. Such additional parameters result in various central extensions of the supersymmetry algebra.
\end{itemize}

One shortcoming of these constructions is that they do not include ${\cal N} = 4$ non-abelian gauge theories with Chern-Simons interactions, such as the ones in \cite{Gaiotto:2008ak,Aharony:2008gk,Aharony:2008ug,Imamura:2008dt,Hosomichi:2008jd}. For such theories there is currently no off-shell description available, and hence they fall outside of the class of theories that can be studied using our methods.

\subsection{$\cN=4$ Weyl multiplet and squashed sphere backgrounds} 

\subsubsection{Weyl multiplet}

$\cN=4$ conformal supergravity is obtained by promoting the 3d $\cN=4$ superconformal symmetry to a local symmetry. The 3d $\cN=4$ superconformal group is $\OSp(4|4)$ whose maximal bosonic subgroup is given by $\SO(4)_R\times \USp(4)$. The associated gauge fields together with some auxiliary fields form the ${\cal N} = 4$ Weyl multiplet \cite{Banerjee:2015uee,Butter:2013rba,Butter:2013goa}.\footnote{We use the conventions of \cite{Banerjee:2015uee}.}  The field content of this multiplet is given by\footnote{In addition there are also gauge fields $f_\mu{}^a$ and $\phi_\mu{}^{ip}$ for special conformal symmetries and supersymmetries. However, these satisfy curvature constraints and can be expressed as composite fields in terms of the other fields in the Weyl multiplet.}
\begin{equation*}
	\begin{aligned}
		{\rm Bosons:}&\quad e_\mu{}^a\,,\quad  b_\mu\,,\quad V_\mu{}^i{}_j \,,\quad \wti V_\mu{}^p{}_q\,,\quad C\,,\quad D\,,\\
		{\rm Fermions:}&\quad \psi_\mu^{ip}\,,\quad \chi^{ip}\,.\\
	\end{aligned}
\end{equation*}
$e_\mu{}^a$ is the vielbein while $b_\mu$, $V_\mu{}^i{}_j$, and $\wti V_\mu{}^p{}_q$ are the gauge fields for dilatations and the $\SU(2)_H$ and $\SU(2)_C$ factors of the $\SO(4)_R\simeq \SU(2)_H\times \SU(2)_C$ R-symmetry, respectively. $\psi_\mu^{ip}$ are the Poincar\'e supersymmetry generators, and in addition there are the auxiliary spinors $\chi^{ip}$ and scalars $C$ and $D$. The indices $\mu,\nu,\dots$ and $a,b,\dots$ are curved and tangent space indices, respectively, while $i,j,\dots$ and $p,q,\dots$ are $\SU(2)_H$ and $\SU(2)_C$ fundamental indices.  More details on our conventions as well as on the Weyl multiplet and its supersymmetry variations can be found in Appendices~\ref{app:conventions} and \ref{app:confSugra}.

When looking for supersymmetric backgrounds, it is convenient to set the values of all fermions to zero, hence the supersymmetry variations of all bosonic fields automatically vanish. In order for the background to preserve supersymmetry, we then need to require the fermionic variations to vanish as well:
\begin{equation}\label{FermVariations}
	\delta \psi_\mu^{ip} = 0\,,\qquad \delta \chi^{ip} = 0 \,.
\end{equation}
The explicit form of these equations is given in Appendix~\ref{app:confSugra}, which we reproduce here for reader's convenience. We have 
\begin{equation}\label{SUSYVarsWeyl}
	\begin{aligned}
		\delta\psi_\mu^{ip} &= 2\cD_\mu \epsilon^{ip} - \gamma_\mu \eta^{ip} = 0\,,\\
		\delta\chi^{ip} &= 2\slashed{D}C \epsilon^{ip} + D\epsilon^{ip} + \f12 \slashed{G}^i{}_j \epsilon^{jp} - \f12 \slashed{\wti G}^p{}_q \epsilon^{iq}  + 2C \eta^{ip} = 0\,,
	\end{aligned}
\end{equation}
where $G$ and $\wti G$ are the field strengths associated to $V$ and $\wti V$, respectively, and $\epsilon^{ip}$ and $\eta^{ip}$ are the parameters for the Poincar\'e and conformal supersymmetry transformations, respectively. $\cD$ is the superconformal covariant derivative which acts on the supersymmetry parameter $\epsilon$ as 
\begin{equation}\label{cDDef}
	\cD_\mu \epsilon^{ip} = \left( \partial_\mu + \f14 \omega_\mu^{ab}\gamma_{ab}+\f12 b_\mu \right)\epsilon^{ip} + \f12 V_\mu{}^i{}_j \epsilon^{jp} + \f12 \wti V_\mu{}^p{}_q \epsilon^{iq}\,.
\end{equation}

We are now ready to describe the backgrounds of interest.  As already mentioned, the metric takes the form \eqref{3dmetric}, for which we choose the following vielbein:
\begin{equation}\label{3dvielbein}
	\begin{aligned}
		e_1 &= - \f r2 \left( \sin\psi\dd\theta - \sin\theta\cos\psi\dd\phi \right)\,,\\
		e_2 &= \f r2 \left( \cos\psi\dd\theta - \sin\theta\sin\psi\dd\phi \right)\,,\\
		e_3 &= - \f r2 \f{b+b^{-1}}{2}\left( \dd\psi + \cos\theta\dd\phi \right)\,. 
	\end{aligned}
\end{equation}
For the remaining fields, we find two backgrounds presented in turn below.

\subsubsection{$\mathfrak{su}(2|1) \times \mathfrak{psu}(1|1)$-invariant background}
\label{FIRSTBACKGROUND}

The first background corresponds to the following values for the Weyl multiplet fields: 
\begin{align}
	V{}^i{}_j &=-\f i2 V (\sigma_3){}^i{}_j\,, &  \wti V^p{}_q &=-\f i2 \wti V (\sigma_3){}^p{}_q\,,\nn\\
	C &= -\f{i}{2r}\left( b-b^{-1} \right)\,, & D &= -\f{b^{2}-b^{-2}}{2r^2}\,,\\
	b_\mu &= 0 \,, & &\nn
\end{align}
where 
\begin{equation}
	V = -\f{2}{b r}\f{b-b^{-1}}{b+b^{-1}}\left( \dd\psi + \cos\theta\dd\phi \right)\,,\qquad \wti V =  \f{2b}{r}\f{b-b^{-1}}{b+b^{-1}}\left( \dd\psi + \cos\theta\dd\phi \right)\,.
\end{equation}
With these choices, Eqs.~\eqref{SUSYVarsWeyl} are obeyed provided that
\begin{equation}\label{epsetaEqs}
	2\cD_a \epsilon^{ip} = \gamma_a \eta^{ip}\,,\qquad \eta^{ip}= \f{i}{r}\left[ \f12 \left(b+b^{-1}\right)\epsilon^{ip} + b^{-1}\gamma_3(\sigma_3)^i{}_j\epsilon^{jp} + b \,\gamma_3 (\sigma_3)^p{}_q\epsilon^{iq} \right] \,.
\end{equation}
When $b=1$, these equations simplify, and one recovers the conformal Killing spinor equation on the round sphere, which has eight linearly-independent solutions. When $b\neq 1$, on the other hand, the equations \eqref{epsetaEqs} have six linearly-independent solutions given by
\begin{equation}\label{supercharges}
	\epsilon^{21}= \xi\,, \qquad \epsilon^{12} = -i\wti{\xi}\,,\qquad \epsilon^{11}=\begin{pmatrix}
		\zeta_1 \\ 0 
	\end{pmatrix}\,,\qquad 
	\epsilon^{22}=\begin{pmatrix}
		0 \\ \zeta_2 
	\end{pmatrix}\,,
\end{equation}
where the spinors $\xi$ and $\wti{\xi}$ are defined as
\begin{equation}\label{genspinor}
	\xi = \begin{pmatrix}
		i\sqrt{b} & 1 \\ 1 &-\f{i}{\sqrt{b}} 
	\end{pmatrix}
	g^{-1}
	\begin{pmatrix}
		\xi_1 \\ \xi_2 
	\end{pmatrix}\,,\qquad 
	\wti{\xi} = \begin{pmatrix}
		-\f{i}{\sqrt{b}} & 1 \\ 1 & i \sqrt{b} 
	\end{pmatrix}
	g^{-1}
	\begin{pmatrix}
		\wti{\xi}_1 \\ \wti{\xi}_2\end{pmatrix} \,, 
\end{equation}
where $g$ parameterizes an $\SU(2)$ element given by
\begin{equation}
	g= \begin{pmatrix}
		\cos\f{\theta}{2}\e^{\f{i}{2}(\phi+\psi)} & \sin\f{\theta}{2}\e^{\f{i}{2}(\phi-\psi)} \\ -\sin\f{\theta}{2}\e^{-\f{i}{2}(\phi-\psi)} & \cos\f{\theta}{2}\e^{-\f{i}{2}(\phi+\psi)} 
	\end{pmatrix}\,.
\end{equation}

In formulating this background we chose a particular embedding of $\UU(1)_H\times \UU(1)_C$ in $SU(2)_H \times SU(2)_C$. However, we could have replaced $(\sigma_3)^i{}_j$ and $(\sigma_3)^p{}_q$ by a different choice of Cartan elements. Furthermore, upon inspection of the solution above, one can see that the $\SU(2)_H$ and $\SU(2)_C$ vector fields appear in a symmetric way. Indeed, analogous to the background introduced above, we can define a mirror dual background by performing the interchange
\begin{equation}\label{MirrorMap}
	V_\mu{}^i{}_j \leftrightarrow \wti V_\mu{}^p{}_q\,,\qquad C \leftrightarrow -C \,,\qquad D \leftrightarrow -D\,.
\end{equation}

Before moving on to the next squashed sphere background let us justify the title of this subsection and show that the supersymmetry transformations of this background generate the superalgebra $\mathfrak{su}(2|1) \times \mathfrak{psu}(1|1)$. A general supercharge is defined by a spinor parameter $\epsilon^{ip}$, as in \eqref{supercharges}, but from this equation it is not immediately clear what the superalgebra is. In order to illustrate this more explicitly, let us introduce the following set of supercharges,
\begin{equation}\label{susies1}
	\begin{aligned}
		Q_1^{(l+)} 	&:\, \epsilon\left( \xi_1 = i \right)\,,\quad 	& 	Q_1^{(l-)} 	&:\,\epsilon\left( \tilde{\xi}_1 = 1 \right) \,,\\
		Q_2^{(l+)} 	&: \epsilon\left( \xi_2 = i \right)\,,\quad 	& 	Q_2^{(l-)} 	&: \epsilon\left( \tilde{\xi}_2 = 1 \right)\,,\\
		Q^{(r+)}  	&: \epsilon\left( \zeta_1 = 1 \right) \,,\quad & 	Q^{(r-)} 	&: \epsilon\left( \zeta_2 = 1 \right) \,,
	\end{aligned}
\end{equation}
where the parameters denoted between the brackets indicate the values of the non-zero parameters in the spinors \eqref{supercharges}--\eqref{genspinor}. As our superalgebra is constructed as a subalgebra of the $\cN=4$ superconformal algebra, we can compute the (anti-)commutation relations using the $\cN=4$ superconformal algebra, given in detail in Appendix~\ref{app:susyAlgebra}\@. Using Eq.~\eqref{genQcomm}, one can straightforwardly show that all the supercharges \eqref{susies1} are all nilpotent. Furthermore, the supercharges with superscript $(l\pm)$, together with the generators $J_{\alpha\beta}^l$ and $R^l$ generate an $\su(2|1)$ algebra with the non-zero (anti-)commutation relations given by
\begin{equation}\label{lalgebra1}
	\begin{aligned}
		\comm{J^l_{i}}{J^l_{j}} &= i\varepsilon_{ijk}J_k^l\,, & \comm{J_{\alpha\beta}^{l}}{Q_\gamma^{(l\pm)}} &= \f{1}{2}\left( \varepsilon_{\alpha\gamma}Q_\beta^{(l\pm)} + \varepsilon_{\beta\gamma}Q_\alpha^{(l\pm)}\right)\,,\\
		\comm{R^l}{Q_\alpha^{(l\pm)}} &= \pm Q_\alpha^{(l\pm)}\,,\qquad& \acomm{Q_\alpha^{(l+)}}{Q_\beta^{(l-)}} &= -\f{4i}{r}\left( J^{l}_{\alpha\beta} + \varepsilon_{\alpha\beta}R^l\right)\,.
	\end{aligned}
\end{equation}
In these equations we introduced the $\SU(2)$-triplet $J_i\equiv - \f12 \varepsilon^{\beta\gamma}(\sigma_i)^\alpha{}_\gamma J_{\alpha\beta}$. Similarly, the supercharges with superscript $(r\pm)$ together with the generators $J^r$ and $R^r$ generate a $\psu(1|1)$ algebra with commutation relations
\begin{equation}\label{lalgebra2}
	\begin{aligned}
		\comm{J^{r}}{Q^{(r\pm)}} &= \pm\f{1}{2} Q^{(r\pm)}\,, &	\comm{R^r}{Q^{(r\pm)}} &= \pm Q^{(r\pm)}\,,\\
		\acomm{Q^{(r+)}}{Q^{(r-)}} &= -\f{4i}{r}\left( J^{r} + R^r\right)\,.\qquad&&
	\end{aligned}
\end{equation}
The generators $J^l_{\alpha\beta}$ and $J^r$ generate the $\su(2)\times \uu(1)$ isometries of the squashed sphere and act on gauge-invariant operators, $\cO$, as
\begin{equation}\label{actionJ}
	J_i^l\cO = -\cL_{v^l_i}\cO\,,\qquad J^r\cO = -\cL_{v^r}\cO\,,
\end{equation}
where $\cL_v$ denotes the Lie derivative with respect to the Killing vector $v$. In our coordinates, the Killing vectors take the form
\begin{align}\label{Killingv}
	v_1^l &= i \Big( \sin\phi\,\partial_\theta + \cos\phi\left( \cot\theta\,\partial_\phi - \csc\theta\,\partial_\psi \right) \Big) \,,\\
	v_2^l &= i \Big( \cos\phi\,\partial_\theta - \sin\phi\left( \cot\theta\,\partial_\phi - \csc\theta\,\partial_\psi \right) \Big) \,,\\
	v_3^l &= i\,\partial_\phi \,,\\ 
	v^r &= \f{2i}{b+b^{-1}}\,\partial_\psi\,.
\end{align}
To define the action of the R-symmetries on the other hand, it is useful to define the Cartan generators of the $\SU(2)_H\times \SU(2)_C$ R-symmetries as
\begin{equation}
	\bH = \f12 (\sigma_3)^i{}_j \bH^j{}_i\,,\qquad \bC = \f12 (\sigma_3)^p{}_q \bC^q{}_p\,,
\end{equation}
whose action on gauge invariant operators is defined in \eqref{Raction}. In terms of these generators, the left and right R-symmetry, $R^l$ and $R^r$, can be written as
\begin{equation}
	R^l = \f{1}{2(b+b^{-1})}\left(b\,\bH - \f 1b \bC \right)\,,\qquad R^r = \f{1}{2(b+b^{-1})}\left(\bH + \bC \right)\,.
\end{equation}
%

%%%%%%%%%%%%%%%%%%%%%%%%%%%%%%%%%%%%%%%%%%%%%%%%%%%%%%%%%%%%%%%%%%%%%%%%%%%%%%%%
\subsubsection{$\mathfrak{psu}(2|2) \times \mathfrak{u}(1)$-invariant background}
\label{BACKGROUND2}
%%%%%%%%%%%%%%%%%%%%%%%%%%%%%%%%%%%%%%%%%%%%%%%%%%%%%%%%%%%%%%%%%%%%%%%%%%%%%%%%

To construct the second background, we start from the the same metric \eqref{3dmetric} and vielbein \eqref{3dvielbein}. However, in this case, we complete the supersymmetric background with different choices for the other background fields in the Weyl multiplet, namely
\begin{align}
	V{}^i{}_j &= 0 \,, & \wti V^p{}_q &=-\f i2 \wti V (\sigma_3){}^p{}_q\,, \nn\\
	C &= -\f{i}{2r}\left( b+b^{-1} \right)\,, & D &= -\f{\left(b+b^{-1}\right)^2}{2r^2}\,, \\
	b_\mu &= 0\,,& &\nn 
\end{align}
where 
\begin{equation}
	\wti V =  \f{2}{r}\left(b-b^{-1}\right)\left( \dd\psi + \cos\theta\dd\phi \right)\,.
\end{equation}
With these choices, all supersymmetry variations \eqref{SUSYVarsWeyl} vanish provided that
\begin{equation}
	2\cD_a \epsilon^{ip} = \gamma_a \eta^{ip}\,,\qquad \eta^{ip}= \f{i}{r}\left[ \f12(b+b^{-1})\epsilon^{ip} + (b-b^{-1})\gamma_3(\sigma_3)^p{}_q\epsilon^{iq} \right]\,.
\end{equation}
When $b=1$ the first term in the expression for $\eta^{ip}$ vanishes, and we again recover the conformal Killing spinor equation on the round sphere. In this case, however, for non-zero $b\neq 1$, these equations still allow for eight independent supercharges given by
\begin{equation}\label{supercharges2}
	\epsilon^{21} = \xi\,, \qquad \epsilon^{12} = -i\wti{\xi}\,,\qquad \epsilon^{11} = \widehat{\xi} \,,\qquad 
	\epsilon^{22}= -i\widehat{\wti\xi} \,,
\end{equation}
where the spinors $\xi$ and $\wti{\xi}$ were defined in \eqref{genspinor}, and, similarly, $\widehat{\xi}$ and $\widehat{\wti\xi}$ are given by the same expressions but with independent constants $\xi_3$, $\xi_4$ and $\wti\xi_3$, $\wti\xi_4$.  

As anticipated in the title of this subsection, the supersymmetry algebra preserved by this background is given by $\psu(2|2) \times \uu(1)$. To clarify the structure of this superalgebra let us again introduce a set of supercharges $Q_\alpha^i$ and $\wti Q^i_\alpha$,
\begin{equation}\label{susies2}
	\begin{aligned}
		Q_1^1 	&:\, \epsilon\left( \xi_1 = i \right)\,,\quad 	& 	Q_1^2	&:\,\epsilon\left( \tilde{\xi}_1 = i \right) \,,\\
		Q_2^1 	&: \epsilon\left( \xi_2 = i \right)\,,\quad 	& 	Q_2^2 	&: \epsilon\left( \tilde{\xi}_2 = i \right)\,,\\
		\wti Q_1^1 	&:\, \epsilon\left( \wti \xi_3 = 1 \right)\,,\quad 	& 	\wti Q_1^2 	&:\,\epsilon\left( \xi_3 = 1 \right) \,,\\
		\wti Q_2^1 	&: \epsilon\left( \wti \xi_4 = 1 \right)\,,\quad 	&  \wti	Q_2^2 	&: \epsilon\left( \xi_4 = 1 \right)\,,
	\end{aligned}
\end{equation}
in an analogous fashion to \eqref{susies1}. Together with the $\su(2) \times \su(2)_H$ generators, $J_{\alpha\beta}$ and $\bH^i{}_j$, these supercharges generate the $\psu(2|2)$ algebra, whose non-zero (anti-)commutation relations are given by
\begin{align}
	\comm{J_{i}}{J_{j}} &= i\varepsilon_{ijk}J_k\,, & \comm{\bH^i{}_j}{\bH^k{}_l} &= \delta^k_j \bH^i{}_l - \delta^i_l \bH^k{}_j\,,\\
	\comm{J_{\alpha\beta}}{Q_\gamma^{i}} &= \f{1}{2}\left( \varepsilon_{\alpha\gamma}Q_\beta^{i} + \varepsilon_{\beta\gamma}Q_\alpha^{i}\right)\,, & \comm{\bH^i{}_j}{Q_\alpha^{k}} &= \delta^k_j Q^i_{\alpha} -\f12 \delta^i_j Q^k_\alpha\,,\\
	\comm{J_{\alpha\beta}}{\wti Q_\gamma^{i}} &= \f{1}{2}\left( \varepsilon_{\alpha\gamma}\wti Q_\beta^{i} + \varepsilon_{\beta\gamma}\wti Q_\alpha^{i}\right)\,, & \comm{\bH^i{}_j}{\wti Q_\alpha^{i}} &= \delta^k_j \wti Q^i_{\alpha} -\f12 \delta^i_j \wti Q^k_\alpha\,,
\end{align}
\begin{equation}
	\acomm{Q_\alpha^{i}}{\wti Q_\beta^{j}} = -\f{4i}{r}\left( \epsilon^{ij} J_{\alpha\beta} + \varepsilon_{\alpha\beta}\varepsilon^{ik} \bH^j{}_k% + \varepsilon^{ij}\varepsilon_{\alpha\beta} Z_3
	\right)\,,
\end{equation}
where the Killing vectors $v_i$ corresponding to $J_i$ are identical to those of the first background and given, together with the additional $\uu(1)$ Killing vector, in \eqref{Killingv}. The action of the $\su(2)\times \su(2)_H\times \uu(1)$ generators on gauge invariant operators is identical as for the previous case and defined in \eqref{actionJ} and \eqref{Raction}, respectively. Similar to the previous case we can also consider the mirror dual background in this case by interchanging the role of $\SU(2)_H$ and $\SU(2)_C$ and performing the mirror map \eqref{MirrorMap} on the background fields in the Weyl multiplet.

\subsection{Matter multiplets}

Having identified two $\cN=4$ supersymmetric squashed sphere backgrounds, the next question is how to write down actions for dynamical fields in these backgrounds.  In order to do so, let us briefly review the multiplets one can use in 3d ${\cal N} = 4$ theories. In this paper we will be interested in theories with vector multiplets, twisted vector multiplets, hypermultiplets, and twisted hypermultiplets.\footnote{In addition, one could add a number of (twisted) half-hypermultiplets but we will not consider this possibility in this work. However, our framework can in principle be extended to include these cases.}

The components of vector and twisted vector multiplets are given by
\begin{center}
	\begin{tabular}{rcc}
		& $\qquad$Vector multiplet $\cV$$\qquad$ & Twisted vector multiplet $\wti\cV$\\
		\Tstrut Bosons:&  $L^p{}_q\,, \quad Y^i{}_j\,,\quad A_\mu\,,$ & $\wti{L}^i{}_j\,, \quad \wti{Y}^p{}_q\,,\quad \wti{A}_\mu\,,$ \\
		Fermions:& $\Omega^{ip}\,$ &  $\wti\Omega^{ip}\,.$
	\end{tabular}
\end{center}
The vector multiplet consists of two triplets of scalars, $L^p{}_q$ and $Y^i{}_j$, a gauge field $A_\mu$, and a gaugino, transforming, respectively, in  the $(\mathbf{1},\mathbf{3})$, $(\mathbf{3},\mathbf{1})$, $(\mathbf{1},\mathbf{1})$, and $(\mathbf{2},\mathbf{2})$ representations of $\SU(2)_H \times \SU(2)_C$.  In the non-abelian case, all these fields transform in the adjoint representation of a classical Lie group.  The twisted vector multiplet has identical field content to the vector multiplet, with the only difference being that the $\SU(2)_H$ indices are interchanged with the $\SU(2)_C$ ones. When continued to Lorentzian signature, the bosonic fields satisfy the following reality conditions 
\begin{equation}
	A_\mu^\dagger = A_\mu\,, \qquad \left( L^p{}_q \right)^\dagger = L^q{}_p\,, \qquad  \left( Y^i{}_j \right)^\dagger = - Y^j{}_i\,,
\end{equation}
as well as
\begin{equation}
	\wti A_\mu^\dagger = \wti A_\mu\,, \qquad \left( \wti L^i{}_j \right)^\dagger = \wti L^j{}_i\,, \qquad  \left(\wti Y^p{}_q \right)^\dagger = -\wti Y^q{}_p\,.
\end{equation}

The components of $n_h$ (ungauged) hypermultiplets and $\tilde n_h$ (ungauged) twisted hypermultiplets are given by:
\begin{center}
	\begin{tabular}{rcc}
		& $\qquad$ hypermultiplets $\cH^I$$\qquad$ & Twisted hypermultiplets $\wti\cH^{\wti I}$\\
		\Tstrut Bosons:&  $z^{iI}\,,$ & $\wti z^{p \tilde I}\,,$ \\
		Fermions:& $\zeta^{Ip}\,$ &  $\wti\zeta^{\tilde I i}\,,$
	\end{tabular}
\end{center}
where the index $I=1,\cdots, 2n_h$ and $\tilde I = 1, \cdots, 2 \tilde n_h$ are fundamental indices of $\USp(2n_h)$ and $\USp(2 \tilde n_h)$, respectively.  In a gauge theory, a subgroup of $\USp(2n_h)$ and/or $\USp(2 \tilde n_h)$ could be gauged under vector multiplets and twisted vector multiplets, respectively. The complex scalars $z^{iI}$ transform as doublets under $\SU(2)_H$, while the fermions $\zeta^{Ip}$ transform as doublets under $\SU(2)_C$.  The fields of the twisted hypermultiplets have similar properties, with $\SU(2)_H$ and $\SU(2)_C$ interchanged.

While the hypermultiplet scalars $z^{iI}$ are individually complex, complex conjugation relates them to one another. When continued to Lorentzian signature, the reality condition is given by
\begin{equation}\label{RealityCond}
	(z^{iI})^* =  \varepsilon_{IJ} z_i{}^J   \,,
\end{equation}
where $\varepsilon_{IJ}$ is an antisymmetric rank-two invariant tensor of $\USp(2n_h)$.  For concreteness, one can take $\varepsilon_{IJ}$ to be of the form
\es{epsIJExplicit}{
	\varepsilon_{IJ} = \begin{pmatrix}
		i \sigma_2 & 0 & 0 & \cdots \\
		0 & i \sigma_2 & 0 & \cdots \\
		0 & 0 & i \sigma_2 & \cdots \\
		\vdots & \vdots & \vdots & \ddots 
	\end{pmatrix} \,.
}
Similarly, the twisted hypermultiplet scalars $\wti z^{p \tilde I}$ obey the reality conditions 
\es{RealityTwisted}{
	(\wti z^{p \tilde I})^* =  \varepsilon_{\tilde I \tilde J} \wti z_p{}^{\tilde J} 
}
when continued to Lorentzian signature.

\subsection{Conformal matter actions}

Having reviewed the multiplets, we are now ready to write down various terms in the action. However, since so far we have discussed only a conformal supergravity background without the addition of compensator multiplets, we are restricted to the following conformal actions: 
\begin{itemize}
	\item Kinetic terms for hypermultiplets coupled to a vector multiplet,
	\item Kinetic terms for twisted hypermultiplets coupled to a twisted vector multiplet,
	\item Mixed abelian Chern-Simons terms.
\end{itemize}

The bosonic part of the kinetic term in the action of the hypermultiplets $\cH^I$ coupled a vector multiplet $\cV$ is given by
\begin{equation}\label{hypLag}
	\begin{aligned}
		S_{\rm hyp}[\cH^I, \cV] = -\f12 \int d^3x\,  \sqrt{g}\,  \varepsilon_{IJ}& \Bigg( 
		\cD_\mu z_i{}^I\cD^\mu z^{iJ} - \f14 z_i{}^I z^{iJ}\left( -\f12 R + D - C^2 \right) \\
		&{}\quad+ \f{1}{2} z_i{}^I L^p{}_q{}^J{}_K L^q{}_p{}^K{}_L z^{iL} + i\,z_i{}^I Y^{i}{}_j{}^J{}_K z^{jK}\\
		&{}\quad+ i \overline\zeta^{pI}\slashed{\cD}\zeta_p^J + i \overline\zeta_p^I L^p{}_q \zeta^{qJ} + z_i^I \overline\Omega^{ip}\zeta_p^J \Bigg)\,,
	\end{aligned}
\end{equation}
where the covariant derivative acting on the hyperscalars is given by
\es{CovDerScalars}{
	\cD_a z_i{}^I \equiv \partial_a z_i{}^I + \f12  z_j{}^I V_a{}^j{}_i - i A_a{}^I{}_J z_i^J\,.
}
In the kinetic term \eqref{hypLag}, we wrote the vector multiplet fields explicitly as $2n_h \times 2n_h$ matrices acting in the representation in which the $z^{iI}$ transform. 

The kinetic term in the action for twisted hypermultiplets coupled to a twisted vector multiplet is analogous as the one for hypermultiplets and can be obtained by interchanging the hyper and vector multiplet fields with their twisted counterparts, while at the same time performing the mirror map \eqref{MirrorMap} on the background fields in the Weyl multiplet.

Finally, the mixed Chern-Simons term between an abelian vector multiplet and an abelian twisted vector multiplet is given by \cite{Kapustin:1999ha}
\begin{equation}\label{SBF}
	S_{\rm BF} \propto
	\f{k}{4\pi}\int A \wedge \dd \wti A + \f{k}{2\pi}\int d^3x\, \sqrt{g}\, i\left( \overline \Omega_{ip} \wti\Omega^{ip}  + \frac{1}{2}  Y^i{}_j \wti L^j{}_i + \frac{1}{2} L^p{}_q \wti Y^q{}_p  \right) \,. 
\end{equation}

\subsection{Non-conformal actions, real masses and FI parameters}\label{subsec:NONCONFORMAL}

So far, we have introduced a collection of conformal actions. However, this will not suffice for our purposes. In the following we will consider dynamical vector multiplets with a Yang-Mills action. This action is non-conformal in three dimensions, and hence in order to construct it we will need to introduce additional background compensator vector and/or twisted vector multiplets. A second application of adding background multiplets is that they allow us to introduce real masses and/or FI terms. The background values for these multiplets depend on the chosen background for the Weyl multiplet, and hence we will again treat the two backgrounds separately. 

\subsubsection{$\mathfrak{su}(2|1) \times \mathfrak{psu}(1|1)$-invariant background}
\label{NONCONFORMAL}

Let us first consider an abelian background vector multiplet, $\cV_\text{bk} = \left\{ L_\text{bk}{}^p{}_q ,\, Y_\text{bk}{}^i{}_j,\, A_\text{bk},\, \Omega_\text{bk}^{ip} \right\}$.  As we will explain now, such a multiplet can be either coupled to a conserved current multiplet made out of the dynamical fields or it can be used to construct a supersymmetric Yang-Mills action.  

In either case, one should give the fields in $\cV_\text{bk}$ supersymmetry-preserving expectation values. As usual, we start by giving vanishing expectation values to the fermions, $\Omega_\text{bk}^{ip} = 0$. In order to ensure that the background is supersymmetric, we then need to require that the supersymmetry variation for the gaugino $\Omega_{\rm bk}^{ip}$ vanishes,\footnote{For a non-abelian vector multiplet this supersymmetry variation should be supplemented with an extra term proportional to the commutator of two $L^p{}_q$s, the full non-abelian supersymmetry variations are given in \eqref{SUSYTvec}.}
\es{VanishFerm}{
	\delta\Omega_\text{bk}{}^{ip} &= \slashed{\cD}L_\text{bk}{}^p{}_q\epsilon^{iq}-\f12 F_{\text{bk}\,ab}\gamma^{ab}\epsilon^{ip}+ Y_\text{bk}{}^i{}_j\epsilon^{jp}+C L_\text{bk}{}^p{}_q\epsilon^{iq} + L_\text{bk}{}^p{}_q\eta^{iq} = 0 \,,
}
where $F_{\rm bk}$ is the field strength associated to the gauge field $A_{\rm bk}$. In the $\mathfrak{su}(2|1) \times \mathfrak{su}(1|1)$ background, this equation is solved for the following values of the background fields in the abelian vector multiplet
\es{VectorBk}{
	L_\text{bk}{}^p{}_q = \f{\mu}{r} \left(\sigma_3\right)^{p}{}_q \,, \qquad
	A_\text{bk} =\f{\mu}{r} \, \f{b^{-1}-b}{b^{-1}+b}\left( \dd\psi + \cos\theta\dd\phi \right)\,,\qquad Y_\text{bk}{}^{i}{}_j = \f{i \mu}{b\,r^2}(\sigma_3)^i{}_j \,,
}
where $\mu$ is an arbitrary parameter. In the non-abelian case, the vector multiplet fields as well as the parameter $\mu$ become Lie algebra-valued. This background preserves all supersymmetries and hence the full $\mathfrak{su}(2|1) \times \mathfrak{psu}(1|1)$ superalgebra. 

Analogously, one can construct a background for an abelian twisted vector multiplet $\wti\cV_\text{bk} = \left\{ \wti L_\text{bk}{}^p{}_q ,\, \wti Y_\text{bk}{}^i{}_j,\, \wti A_\text{bk},\, \wti\Omega_\text{bk}^{ip} \right\}$ that again preserves the full $\mathfrak{su}(2|1) \times \mathfrak{psu}(1|1)$ algebra. The supersymmetry preserving background values for the fields in the twisted multiplet are given by
\es{TwistedVectorBk}{
	\wti{L}_\text{bk}{}^i{}_j = \f{\wti \mu}{r}\left(\sigma_3\right)^{i}{}_j \,, \qquad
	\wti{A}_\text{bk} = \f{\wti \mu}{r}\,\f{b-b^{-1}}{b+b^{-1}}\left( \dd\psi + \cos\theta\dd\phi \right)\,,\qquad& \wti{Y}_\text{bk}{}^{p}{}_q = \f{i \, b\,\wti \mu }{r^2}(\sigma_3)^p{}_q\,,
}
for an arbitrary parameter $\wti \mu$.  In the non-abelian case, the parameter $\wti \mu$ as well as the twisted vector multiplet fields are Lie algebra-valued.

A first application of such background (twisted) vector multiplets is that they allow us to construct the Yang-Mills action for dynamical vector multiplets ${\cal V}$ or twisted vector multiplet $\wti \cV$. The construction of the Yang-Mills action for a dynamical vector multiplet in a general $\cN=4$ supersymmetric background is discussed in more detail in Appendix~\ref{LAGRANGIANS} and crucially involves coupling the Weyl multiplet to an abelian compensating vector multiplet ${\cal V}_0$. This coupling gauge fixes part of the conformal symmetries and hence allows us to consider non-conformal actions. Specifying to the $\mathfrak{psu}(2|1) \times \mathfrak{su}(1|1)$-invariant squashed sphere background, we can put the background values of the compensating vector multipet to be ${\cal V}_0 = {\cal V}_\text{bk}$ as given in \eqref{VectorBk}. After an appropriate rescaling, one then obtains the following Yang-Mills action,
\begin{equation}\label{YMTerm}
	\begin{aligned}
		S_{\rm YM}[\cV] =&{} \frac{1}{g_\text{YM}^2} \int d^3 x\,  \sqrt{g} \, \bigg\{   \cD_\mu L^p{}_q\cD^\mu L^q{}_p + F_{\mu\nu}F^{\mu\nu} - Y^i{}_j Y^j{}_i \\
		&{}+ \f{3b^2-b^4-1}{b^2 r^2}L^p{}_qL^q{}_p - \left(b-b^{-1}\right)\frac{3-2b^2}{2br^2}\left[(\sigma_3)^p{}_qL^q{}_p\right]^2\\  
		&{}+ \f{i}{b r}(\sigma_3)^j{}_i(\sigma_3)^q{}_p L^p{}_q Y^i{}_j - \f{1}{4}\comm{L^p{}_q}{L^r{}_s}\comm{L^q{}_p}{L^s{}_r}\\
		&{}+ i\overline\Omega^{ip}\slashed{\cD}\Omega_{ip} + \f{1}{2b r}(\sigma_3)^j{}_i(\sigma_3)^q{}_p\overline\Omega^{ip}\Omega_{jq} - \overline\Omega^{ip}\comm{\Omega_{iq}}{L^q{}_p}    \bigg\}\,.
	\end{aligned}
\end{equation}
Similarly, by coupling the Weyl multiplet to an abelian compensator twisted vector multiplet one can construct the Yang-Mills action for a twisted vector multiplet. The resulting action is related to the Yang-Mills action above through mirror symmetry.

The second application involving background vector and twisted vector multiplets is to introduce real masses and Fayet-Iliopolous terms that can be obtained by coupling these multiplets to conserved current multiplets. Just as there are two types of vector multiplets (vectors and twisted vectors) in 3d ${\cal N} = 4$ theories there are two types of conserved current multiplets, namely conserved current multiplets $\cJ=(J^i{}_j, \Xi^{ip}, j_a, K^p{}_q)$ that couple to vector multiplets, and twisted conserved current multiplets $\wti \cJ=(\wti J^i{}_j, \wti \Xi^{ip}, \wti j_a, \wti K^p{}_q)$ that couple to twisted vector multiplets. An example of a conserved current multiplet $\cJ$ can be obtained from an abelian twisted vector multiplet $\wti \cV$ via
\es{Ident}{
	J^i{}_j = \wti L^i{}_j \,, \qquad \Xi^{ip} = \wti \Omega^{ip} \,, \qquad j_a = \frac 12 \epsilon_{abc} \wti F^{bc} \,, \qquad K^p{}_q = \wti Y^p{}_q \,,
}
with $\wti F$ the field strength associated to the gauge field $\wti A$. From this equation we see that the conserved current is simply the Hodge dual of the field strength of a twisted vector. For this reason we will sometimes denote such conserved current multiplet as $\cJ = *\wti \cV$. The relations in \eqref{Ident} can be used to determine the supersymmetry transformation rules of the current multiplet from those of the twisted vector multiplet  \eqref{SUSYTtvec} as given in \eqref{currentSUSY}. From these transformation rules, one can deduce that the supersymmetric coupling of a general conserved current multiplet $\cJ$ to a vector multiplet $\cV$ takes the form
\begin{equation}\label{LVJ}
	S_\text{current} [\cV, \cJ] =
	\int d^3x\, \sqrt{g}\, \left[ A_\mu j^\mu + i \overline \Omega_{ip} \Xi^{ip}  + \frac{i}{2}  Y^i{}_j J^j{}_i + \frac{i}{2} L^p{}_q K^q{}_p  \right] \,.
\end{equation}
In the non-abelian case, the only modification to \eqref{LVJ} is that one should take the trace of the expression in the square bracket.  Note that \eqref{LVJ} is already included in the kinetic term for the hypermultiplets in \eqref{hypLag} when one expands the latter to linear order in the vector multiplet fields.  In this case, quadratic terms in the vector multiplet fields are also required in order to preserve supersymmetry. Again, we can write analogous expressions for twisted background vector multiplets. Similar to \eqref{Ident}, the twisted conserved current multiplet, $\wti \cJ = \star \cV$, is given by
\es{Identtwisted}{
	\wti J^p{}_q =  L^p{}_q \,, \qquad \wti\Xi^{ip} = \Omega^{ip} \,, \qquad \wti j_a = \frac 12 \epsilon_{abc} F(A)^{bc} \,, \qquad \wti K^i{}_j =  Y^i{}_j \,,
}
and the coupling to the twisted vector multiplet takes the form analogous to \eqref{LVJ},
\begin{equation}\label{LVJtwisted}
	S_\text{twisted current} [\wti\cV, \wti\cJ] =
	\int d^3x\, \sqrt{g}\, \left[ \wti A_\mu \wti j^\mu + i \overline{\wti\Omega}_{ip} \wti\Xi^{ip}  + \frac{i}{2}  \wti Y^p{}_q \wti J^q{}_p + \frac{i}{2} \wti L^i{}_j \wti K^j{}_i  \right] \,.
\end{equation}
In the presence of flavor symmetries, with global symmetry group $G_F$ acting on the hypermultiplets, one can introduce real mass parameters $m$ valued in the Cartan of the Lie algebra of $G_F$ by coupling the hypermultiplets to a background vector multiplet with background values \eqref{VectorBk} with $\mu = m$.  In other words, instead of considering $S_\text{hyp}[\cH^I, \cV]$, where $\cV$ is a dynamical vector multiplet, one considers 
\es{KinMod}{
	S_\text{hyp}\left[\cH^I, \cV + \cV_\text{bk} \big|_{\mu = m} \right]  \,.
}
Similarly, for every $\UU(1)$ factor in the dynamical gauge group $G$ one can introduce Fayet-Iliopolous parameters $\zeta$ corresponding to background twisted vector multiplets in the Cartan of the gauge group taking the values \eqref{TwistedVectorBk} with $\wti \mu = \zeta$.  Explicitly, for an abelian multiplet $\cV$, the FI term in the action is
\es{FITerm}{
	S_\text{FI}[\cV] =& S_\text{twisted current}[\wti \cV_\text{bk} \big|_{\wti \mu = \zeta}, *\cV]\\
	=& \sum_{k=1}^{\# \UU(1)\text{'s in } G}\zeta_k \int \dd^3 x\sqrt{g}\left( \f{b-b^{-1}}{b+b^{-1}}\,\wti j^3 - \f{b}{2r}(\sigma_3)^p{}_q \wti J^q{}_p + \f{i}{2}(\sigma_3)^i{}_j \wti K^j{}_i \right)
}
Note that, in contrast to theories on flat space, where the FI parameters transform in the $(\mathbf{3},\mathbf{1})$ of $\SU(2)_H\times \SU(2)_C$, on the squashed sphere only the component $\zeta = (\sigma_3)^i{}_j \zeta^j{}_i$, corresponding to the choice of the $\SU(2)_H$ Cartan preserved by the background, can be turned on. Analogously, only a single $\SU(2)_H\times \SU(2)_C$ component of the real mass terms, $m =(\sigma_3)^p{}_q m^q{}_p$ can be turned on on the squashed sphere. 

Once more, one can construct a twisted analog to the previous discussion. When the theory contains flavor symmetries $\wti G_F$ acting on the twisted hypermultiplets, one can introduce real mass terms $\wti m$ valued in the Cartan of the Lie algebra of $\wti G_F$ and for every abelian factor in dynamical twisted gauge group $G$ one can introduce twisted FI parameters $\wti \zeta$ corresponding to background vector multiplets in the Cartan of the twisted gauge group. The couplings are identical to the above and can be obtained by interchanging the roles of the vector and twisted vector multiplets and those of conserved current and twisted conserved current multiplets.

Introducing background (twisted) vector multiplets preserves the full supersymmetry algebra. However, coupling them to (twisted) conserved current multiplets introduces additional central charges in the algebra. From the Jacobi identity, it is clear that the only central charges one can add are given by
\begin{equation}\label{extended1}
	\begin{aligned}
		\acomm{Q_\alpha^{(l+)}}{Q_\beta^{(l-)}} &= -\f{4i}{r}\left( J^{l}_{\alpha\beta} + \varepsilon_{\alpha\beta}R^l + \varepsilon_{\alpha\beta}Z^l \right)\,,\\
		\acomm{Q^{(r+)}}{Q^{(r-)}} &= -\f{4i}{r}\left( J^{r} - R_r - Z_r\right)\,.
	\end{aligned}
\end{equation}
All the other commutators in \eqref{lalgebra1} and \eqref{lalgebra2} remain identical and hence the resulting centrally extended algebra is given by $\left(\su(2|1)\ltimes \bR\right) \times \su(1|1)$, with central charges $Z^l$ and $Z^r$. As shown in Appendix~\ref{app:susyAlgebra}, adding background vector multiplets modifies the supersymmetry algebra. In line with the central extensions, the supersymmetry algebra acquires an additional gauge transformation with gauge parameter $\lambda_{\bQ}$ (see \eqref{QcommQ}). A simple computation using the expression of $\lambda_{\bQ}$ and $\lambda_{\wti \bQ}$ shows that the central charges appearing in \eqref{extended1} are related to the real masses and FI parameters as follows,
\begin{equation}
	\begin{aligned}
		Z^l &= \f{i}{2(b+b^{-1})}\left( b\,m -\f 1b \zeta \right)\,,\\
		Z^r &= \f{i}{2(b+b^{-1})}\left( m + \zeta \right)\,. 
	\end{aligned}
\end{equation}
In these formulae, $m = m_a T_F^a$ is valued in the Cartan of the Lie algebra of the flavor symmetry, where $T_F^a$ are the generators of $G_F$ acting in the appropriate representation. The FI parameter $\zeta = \sum_a \zeta_a t^a$ acts non-trivially only on operators charged under the topological symmetry, where $\zeta_a$ are the FI parameters and the $t^a$ represent the appropriate topological charges. Similarly, when the theory contains twisted hypermultiplets and twisted vector multiplets, an analogous modification of the supersymmetry algebra takes place with parameter $\lambda_{\wti \bQ}$. In this case the central charges are given by
\begin{equation}
	\begin{aligned}
		Z^l &= \f{i}{2(b+b^{-1})}\left( b\,\wti\zeta -\f 1b \wti{m} \right)\,,\\
		Z^r &= \f{i}{2(b+b^{-1})}\left( \wti\zeta + \wti{m} \right)\,,
	\end{aligned}
\end{equation}
where in this case $\wti m =\wti m_a \wti T_{F}^a$ is valued in the Cartan of the twisted flavor symmetry group $\wti G_F$, and the FI parameters are given by $\wti\zeta = \sum \wti\zeta_a \wti t^a$, where $\wti t_a$ are the twisted topological charges for the twisted topological symmetry. When both twisted and regular background multiplets are present, the central charges $Z^l$ and $Z^r$ are simply given by the sums of the two expressions above.

\subsubsection{$\mathfrak{psu}(2|2) \times \mathfrak{u}(1)$-invariant background}

In the $\mathfrak{psu}(2|2) \times \mathfrak{u}(1)$-preserving squashed sphere background presented in Section~\ref{BACKGROUND2}, one cannot find any supersymmetry-preserving configuration for a background twisted vector multiplet.  For a background vector multiplet $\cV_\text{bk} = \{ L_\text{bk}{}^p{}_q ,\, Y_\text{bk}{}^i{}_j,\, A_\text{bk},\, \Omega_\text{bk}^{ip} \}$ on the other hand, the following configuration preserves the full supersymmetry algebra
\begin{equation}
	L_\text{bk}{}^p{}_q = \f{\mu}{r} (\sigma_3)^p{}_q\,,\qquad A_\text{bk} = \f{\mu}{r}  \f{b-b^{-1}}{b+b^{-1}}\left( \dd\psi + \cos\theta\dd\phi \right)\,,\qquad Y_\text{bk}{}^{i}{}_j = 0\,,\label{SUSYvec2}
\end{equation}
where $\mu$ is an arbitrary parameter valued in the Lie algebra of the flavor symmetry. Analogous to the previous background, we can use this background vector multiplet to derive a Yang-Mills action for the dynamical gauge fields or to add real masses for the flavor symmetries $G_F$ acting on the hypermultiplets. In this case it will, however, not be possible to add Yang-Mills terms for dynamical twisted vector multiplets, nor will we be able to add FI terms. 

Using the general formulae derived in Appendix~\ref{LAGRANGIANS}, we can construct a supersymmetric Yang-Mills action for dynamical vector multiplets in this background. The bosonic part of the abelian Yang-Mills action in this case is given by
\begin{equation}\label{YMTerm2}
	\begin{aligned}
		S_{\rm YM}[\cV] =&{} \frac{1}{g_\text{YM}^2} \int d^3 x\,  \sqrt{g} \, \bigg\{   \cD_\mu L^p{}_q\cD^\mu L^q{}_p + F_{\mu\nu}F^{\mu\nu} - Y^i{}_j Y^j{}_i \\
		&{}+ \f{(b-b^{-1})^2}{r^2}\left(-L^p{}_qL^q{}_p +\left[(\sigma_3)^p{}_qL^q{}_p\right]^2\right) - \f{1}{4}\comm{L^p{}_q}{L^r{}_s}\comm{L^q{}_p}{L^s{}_r}\\
		&{}+ i\overline\Omega^{ip}\slashed{\cD}\Omega_{ip} - \overline\Omega^{ip}\comm{\Omega_{iq}}{L^q{}_p}    \bigg\}\,.
	\end{aligned}
\end{equation}
In the presence of flavor symmetries $G_F$, acting on the hypermultiplets, we can again add real masses $m$ by coupling our background to a background vector multiplet valued in the Cartan of the Lie algebra of the $G_F$, with $\mu=m$. These real masses again manifest themselves as central extensions of the supersymmetry algebra. From the Jacobi identity we find that the possible central charges can only appear in the $\acomm{Q}{Q}$ commutators and are given by
\begin{align}
	\acomm{Q_\alpha^{i}}{Q_\beta^{j}} &= \varepsilon_{\alpha\beta}\varepsilon^{ij} Z_1 \,, & \acomm{\wti Q_\alpha^{i}}{\wti Q_\beta^{j}} &= \varepsilon_{\alpha\beta}\varepsilon^{ij} Z_2 \,,
\end{align}
\begin{equation}
	\acomm{Q_\alpha^{i}}{\wti Q_\beta^{j}} = -\f{4i}{r}\left( \epsilon^{ij} J_{\alpha\beta} + \varepsilon_{\alpha\beta}\varepsilon^{ik} \bH^j{}_k + \varepsilon^{ij}\varepsilon_{\alpha\beta} Z_3 \right)\,,
\end{equation}
where the other (anti-)commutators remain identical as before. The resulting supersymmetry algebra is given by $\left(\psu(2|2)\ltimes \bR^3\right)\times \uu(1)$.\footnote{This enlarged $\psu(2|2)\ltimes \bR^3$ algebra is a contraction of the exceptional superalgebra $D(2,1;\alpha)$, with $\alpha\rightarrow 0$ \cite{Beisert:2005tm}. The triplet of central charges is the contraction of the additional $\su(2)$ factor.} Using the expression of $\lambda_{\bQ}$ in \eqref{QcommQ} we can relate these central charges to the real mass parameter $m$ as follows:
\begin{equation}
	Z_1 =  0\,,\qquad Z_2 =  0\,, \qquad Z_3 = m\,.
\end{equation}
As before, the real mass parameter $m=m_a T_F^a$  takes value in the Cartan of the Lie algebra of the flavor symmetry. Therefore, we see that the real mass corresponds to turning on one of the three central charges and the resulting centrally extended superalgebra is given by $\su(2|2)\times \uu(1)$. It would be interesting to better understand what the additional central charges represent in the QFT and if they can be turned on by further deforming the theory.

%%%%%%%%%%%%%%%%%%%%%%%%%%%%%
\section{One-dimensional sectors from cohomology}
\label{sec:1dSectorCohomology}

In the next two sections, we will focus on the centrally-extended $\left(\su(2|1)\ltimes \bR\right) \times \su(1|1)$-invariant squashed sphere. Having described the $\cN=4$ squashed sphere background and its preserved supersymmetry algebra, we now look for protected 1d sectors, similar to those on the round sphere \cite{Dedushenko:2016jxl}. We will find two such theories, where one is related to the $\uu(1)_H$ R-symmetry and will therefore be denoted by $\cT_H$ while the other 1d theory is related to $\uu(1)_C$ R-symmetry and will be denoted by $\cT_C$. Similar questions could be asked for the $\psu(2|2) \times \uu(1)$ background, however, as we show in Appendix \ref{app:anotherOne} in this case the analogous 1d theories do not contain any local operators.

For the round sphere, a cohomological construction of these sectors was given in \cite{Dedushenko:2016jxl}. Even though their results were phrased in terms of a centrally extended $\wti{\su(2|1)} \times \wti{\su(2|1)}$ supersymmetry algebra, a closer examination reveals that the construction of each protected sector relies only on the existence of a particular $\su(1|1)$ subalgebra, which is not contained in any $\cN=2$ subalgebra.\footnote{A similar situation occurs in 4d, where the construction of the VOA relies entirely on the existence of a particular $\su(1|1)$ subalgebra within the $\cN=2$ supersymmetry algebra \cite{Dedushenko:2019yiw,Pan:2019bor}.} Such an $\su(1|1)$ subalgebra exists on the squashed sphere too, so we can proceed analogously to the round sphere case. The following supercharges inside our (centrally extended) $\left(\su(2|1)\ltimes \uu(1)\right)\times \su(1|1)$ superalgebra are associated to the 1d $\cT_H$ theory:
\begin{align}
	\cQ_1^H &= Q_1^{(l+)} - \sqrt{b}Q^{(r-)}\,,& \cQ_2^H &= Q_2^{(l-)} + \sqrt{b}Q^{(r+)}\,,
\end{align}
while the following ones are associated to the 1d $\cT_C$ theory:
\begin{align}
	\cQ_1^C &= Q_1^{(l-)} - \f{1}{\sqrt{b}}Q^{(r-)}\,,& \cQ_2^C &= Q_2^{(l+)} +  \f{1}{\sqrt{b}}Q^{(r+)}\,.
\end{align}

Let us start by focusing on theories built exclusively out of vector and hypermultiplets. As we will see momentarily, in these cases the 1d theories will be related to the Higgs and Coulomb branches of the theory, respectively. Individually, each of these supercharges is nilpotent, but their sums, $\cQ^H=\cQ_1^H+\cQ_2^H$ and $\cQ^C=\cQ_1^C+\cQ_2^C$, satisfy the following non-trivial anti-commutation relations
\begin{align}
	\acomm{\cQ^H}{\cQ^H} &= \f{8}{r}\left( i\,\bP_\beta + i\,\bC + b\,\zeta\,r\right)\,\label{cQH2} \,, \\
	\acomm{\cQ^C}{\cQ^C} &= \f{8}{r}\left( i\,\bP_\beta + i\,\bH + \f{m\,r}{b}\right)\,.\label{cQC2}
\end{align}
In these equations, $\bP_\beta = - i \partial_\beta$ denotes a translation in the angle $\beta=\f12 \left(\psi-\phi\right)$.  In the following discussion it will also be useful to introduce the translation $\bP_\alpha = - i \partial_\alpha$ in the angle $\alpha=\f12 \left(\psi+\phi\right)$.  Note that from the metric \eqref{3dmetric}, the squared norms of the vectors $\partial_\alpha$ and $\partial_\beta$ are
\es{VectorNorms}{
	\lVert \partial_\alpha \rVert^2 &= \lVert \partial_\psi + \partial_\phi \rVert^2 = \frac{r^2 \cos^2 \frac{\theta}{2} ( 1 + 6b^2 + b^4 + (b^2-1)^2 \cos \theta)}{8 b^2} \,, \\
	\lVert \partial_\beta \rVert^2 &= \lVert \partial_\psi - \partial_\phi \rVert^2 = \frac{r^2 \sin^2 \frac{\theta}{2} ( 1 + 6b^2 + b^4 - (b^2-1)^2 \cos \theta)}{8 b^2} \,,
}
and therefore the circle parameterized by $\alpha$ shrinks at $\theta = \pi$, while the one parameterized by $\beta$ shrinks at $\theta = 0$.

A local operator $\cO$ belongs to the equivariant cohomology of $\cQ^{H}$ or $\cQ^C$ if it is annihilated by the right-hand side of \eqref{cQH2} or \eqref{cQC2},
\begin{equation}
	\left(\cQ^{H/C}\right)^2\cO = 0\,.
\end{equation}
Hence, we see that such operators should be invariant under the translation $\bP_\beta$ or, in other words, they should be inserted along the circle at $\theta=0$ parameterized by the angle $\alpha$. In addition, they should be invariant under $\bC$ and $\zeta$ or $\bH$ and $m$, respectively. Next, we can define the twisted translations
\begin{equation}\label{twistedT}
	\begin{aligned}
		\widehat{\bP}_H &= \bP_\alpha + \bH 
		\,,\\
		\widehat{\bP}_C &= \bP_\alpha + \bC 
		\,,
	\end{aligned}
\end{equation}
and note that the left-hand side of the following expressions,
\begin{equation}
	\begin{aligned}
	i\widehat{\bP}_H + \f{m}{r b} &= \f{1}{4}\acomm{\cQ^H}{-Q_2^{(l-)}+\f{1}{b^{3/2}}Q^{(r+)}} \,,\\
	i\widehat{\bP}_C + \f{b\, \wti{m}}{r}  &= \f{1}{4}\acomm{\cQ^C}{-Q_1^{(l-)}+b^{3/2}Q^{(r-)}} \,,
	\end{aligned}
\end{equation}
are $\cQ^H$- and $\cQ^C$-exact, respectively. By the Jacobi identity, it follows that the twisted translations are closed with respect to the supercharges $\cQ^H$ and $\cQ^C$, respectively, and hence can be used to translate cohomology classes along the circle parameterized by $\alpha$. To characterize the equivariant cohomology we can therefore restrict ourselves to operators inserted at $\alpha=0$ and translate them using the twisted translation \eqref{twistedT} to obtain the cohomology classes at $\alpha \neq 0$. Therefore, the cohomology classes of both $\cQ^H$ and $\cQ^C$ form two distinct 1d theories that we will denote by $\cT_H$ and $\cT_C$. Furthermore, whenever $m$ or $\zeta$ vanishes on some operators, the twisted translation on these operators is $\cQ^H$- or $\cQ^C$-exact and, therefore, the correlation functions between these operators will be topological. In this case the OPE is independent of the distance between the operators, but crucially it can depend on the ordering along the circle parameterized by $\alpha$.  For operators which have non-zero eigenvalues for $m$ and $\zeta$, respectively, the correlation functions are no longer topological, but as we will see below, the $\alpha$-dependence remains very simple.

The discussion above shows that any local operators in the equivariant cohomology of $\cQ^H$ or in that of $\cQ^C$ must be inserted at $\theta = 0$. However, we have not shown whether there are any non-trivial operators in these cohomologies, which is a question that we now turn to.

\subsection{Local operators in the  $\cQ^H$-cohomology}
\label{QHCOHOMOLOGY}

As already suggested by the choice of superscript $H$, the 1d $\cT_H$ theory containing the operators in the $\cQ^H$-cohomology are related to the Higgs branch of gauge theories with hypermultiplets and vector multiplets. As a first example, let us consider a theory of free hypermultiplets $\cH^I$, possibly coupled to non-trivial background gauge fields. From the supersymmetry transformations, summarized in Appendix~\ref{app:confSugra}, it is now easy to see that the following linear combination, inserted at $\theta = \alpha=0$,
\begin{equation}
	\cZ^I(0) \equiv \f{1}{\sqrt{2}}\left(z_1^I(0) +  z_2^I(0) \right)\,,
\end{equation}
is $\cQ^H$-invariant. Translating this operator using the twisted translation defined above we find that the operator
\es{calZDef}{
	\cZ^I(\alpha) = \f{1}{\sqrt{2}}\left(\e^{\f i2 \alpha} z_1^I(\alpha) +  \e^{- \f i2 \alpha} z_2^I(\alpha) \right)\,,
}
inserted on the $\theta = 0$ circle parameterized by $\alpha$, is also invariant under $\cQ_H$.  One can then consider polynomials in  $\cZ^I(\alpha)$, which are also $\cQ^H$-invariant.

More generally, if we consider a gauge theory where part of the $\USp(2n_h)$ flavor symmetry of the free hypermultiplets is gauged, the $\cQ^H$-invariant operators are still polynomials in the $\cZ^I$, but one should only restrict them to gauge-invariant polynomials only. In general, it is non-trivial to argue that such operators are not $\cQ^H$-exact, but this fact will become clear in the next section where we compute their correlation functions.

\subsection{Local operators in the $\cQ^C$-cohomology}

In this paper we are mostly interested in studying the $\cT_H$ theory. However, let us briefly give some details on the local operators in the $\cQ^C$-cohomology composing the $\cT_C$ theory. As the subscript $C$ suggest, this sector is related to the Coulomb branch of a theory with vector and hypermultiplets. We can proceed analogous as for the $\cT_H$ theory and observe, using the supersymmetry variations for the vector multiplet, that the following linear combination, inserted at $\theta=\alpha=0$ is $\cQ^C$-invariant:
\begin{equation}
	\cL(0) = \f12\left(L^1{}_1(0)-L^2{}_2(0) + L^1{}_2(0) -  L^1{}_2(0)\right)\,.
\end{equation}
The corresponding twisted translated operator is given by,
\begin{equation}
	\cL(\alpha) = \f12\left( L^1{}_1(\alpha)-L^2{}_2(\alpha) + \e^{\f i2 \alpha}L^1{}_2(\alpha) - \e^{-\f i2 \alpha}L^2{}_1(\alpha) \right)\,.
\end{equation}
For these operators one can straightforwardly extend the analysis we will perform in the next sections and compute their correlation function using the 1d theory $\cT_C$. This is, however, not the whole story for the Coulomb branch as apart from the vector multiplet scalars, the Coulomb branch chiral ring also contains monopole operators that contribute to the 1d protected algebra. In addition, this sector contains a variety of line defect operators called vortex loops, which through mirror symmetry are related to Wilson loops. In this work we will not further consider the $\cT_C$ theory but instead focus on the 1d $\cT_H$ theory. The analogous 1d $\cT_C$ theory on the round sphere was studied in detail in \cite{Dedushenko:2018icp,Dedushenko:2017avn} and we expect a similar story to survive on the squashed sphere. 

\subsection{Gauge theories with twisted multiplets}

In this section, we have so far mainly dealt with gauge theories built out of vector and hypermultiplets. However, as discussed in the previous section such multiplets have twisted analogs for which we can go through the exact same analysis. Therefore, before we continue to study the 1d $\cT_H$ theory for theories with vector and hypermultiplets in more detail, let us briefly comment on their twisted analogs. 

As explained above, the 1d $\cT_H$ theory is related to the $\uu(1)_H$ R-symmetry while the 1d theory $\cT_C$ is related to the $\uu(1)_C$ R-symmetry. For theories built out of regular multiplets, these subscripts are conveniently chosen since the 1d sectors provide respectively a deformation quantization of the Higgs branch and Coulomb branch chiral ring of the 3d theory. For the twisted multiplets on the other hand, the theory $\cT_C$ is related to the twisted Higgs branch, while the theory $\cT_H$ is related to the twisted Coulomb branch.

To illustrate this let us consider a theory of free twisted hypermultiplets $\wti\cH^{\wti I}$, possibly coupled to non-trivial background twisted gauge fields. In this case, it follows from the supersymmetry variations for the twisted hypermultiplet that the following linear combination of twisted hyperscalars, inserted at $\theta=0$ along the circle parameterized by $\alpha$, is $\cQ^C$-invariant,
\begin{equation}\label{tZ}
	\wti \cZ^{\wti I}(\alpha) = \f{1}{2}\left( \e^{\f{i}{2}\alpha}\wti z_1^{\wti I}(\alpha) + \e^{-\f{i}{2}\alpha}\wti z_2^{\wti I}(\alpha) \right)\,.
\end{equation}
Hence, in this case the theory $\cT_C$ consist of the twisted translated fields $\wti \cZ^{\wti I}$ and polynomials thereof. When part of the $\USp(2\wti n_h)$ flavor symmetry of the twisted hypermultiplets is gauged, these polynomials are furthermore constrained to be gauge invariant. 

Analogously, in theories with twisted vector multiplets we can introduce the following twisted translated operators,
\begin{equation}
	\wti\cL(\alpha) = \f12\left( \wti L^1{}_1(\alpha) - \wti L^2{}_2(\alpha) + \e^{\f i2 \alpha} \wti L^1{}_2(\alpha) - \e^{-\f i2 \alpha} \wti L^2{}_1(\alpha) \right)\,,
\end{equation}
that are $\cQ_H$-invariant and thus constitute (part of) the 1d $\cT_H$ theory. For theories built out of twisted vector and hypermultiplets we thus find that the theory $\cT_C$ is very similar to the $\cT_H$ theory for theories with untwisted multiplets. Indeed, in the next section we will show that the results for the twisted Higgs branch theory are analogous to the regular Higgs branch theory and we will explicitly compute the correlation functions of operators of the form \eqref{tZ}. Similar to the previous subsector, we find that the twisted vector multiplet scalar constitutes part of the theory $\cT_H$, but as in the untwisted case this is not the full story, and in order to describe the full 1d theory one has to include monopole operators.

%%%%%%%%%%%%%%%%%%%%%%%%%%%%%%%%%%
\section{The $\cT_H$ sector of gauge theories}
\label{sec:1dSector}
%%%%%%%%%%%%%%%%%%%%%%%%%%%%%%%%%%

Having described the local operators in the 1d sectors, we continue in this section with a more detailed study of the $\cT_H$ sector. In particular, we will derive explicit formulae for computing correlation functions in the 1d Higgs branch sector of 3d gauge theories with vector multiplets and hypermultiplets.  We will also comment on the inclusion of twisted vector multiplets and twisted hypermultiplets.

\subsection{Free massive hypermultiplets}

Let us begin with a single hypermultiplet $\cH^I$, $I = 1, 2$ of mass $m$.  The mass is obtained by coupling the hypermultiplet to a background vector multiplet $\cV_{\text{bk}}$ taking the values in \eqref{VectorBk}, with  $\mu^I{}_J = m (\sigma_3)^I{}_J$ chosen in the $\sigma_3$ direction without loss of generality.  The action is just \eqref{hypLag}:
\es{ActionMassive}{
	S_\text{hyp}[\cH^I, \cV_\text{bk} \big|_{\mu^I{}_J = m (\sigma_3)^I{}_J}] \,.
}
As explained in Section~\ref{QHCOHOMOLOGY}, the operators in the 1d theory are products of the $\cZ^I(\alpha)$ defined in \eqref{calZDef}.  Since the 3d theory is quadratic, it should therefore be possible to calculate the correlators of $\cZ^I(\alpha)$ (or of products thereof) using Wick contractions.  To do that, one would first need to determine the two point function 
\es{TwoPoint}{
	G^{iI}{}_{jJ}(x, x') \equiv \langle z^{iI}(x) z_{jJ}(x') \rangle \,,
}
where $x= (\theta, \phi, \psi)$ and $x'= (\theta', \phi', \psi')$ are two points on the squashed sphere, and then set $\theta = \theta' = 0$ and contract \eqref{TwoPoint} with the functions of position in \eqref{calZDef} to determine the two point function in the 1d theory:
\es{GDef}{
	\cG^I{}_J(\alpha_{12}) = \langle \cZ^I(\alpha_1) \cZ_J(\alpha_2) \rangle = u_i(\alpha_1) u^j(\alpha_2) \cG^{iI}{}_{jJ}(\alpha_1, \alpha_2) \,, \quad
	u^i(\alpha) \equiv \frac{1}{\sqrt{2}} \begin{pmatrix} 
		e^{\frac{i \alpha}{2}} \\
		e^{-\frac{i \alpha}{2}} 
	\end{pmatrix} \,,
}
where   by $\cG^{iI}{}_{jJ}(\alpha_{12})$ we mean $\cG^{iI}{}_{jJ}$ evaluated at two points with $\theta = 0$ and angular coordinates $\alpha_1$ and $\alpha_2$, respectively, and  $\alpha_{12} \equiv \alpha_1 - \alpha_2$.

In principle, the two-point function $G^{iI}{}_{jJ}(x, x')$ can be computed by inverting the kinetic operator in the action, as follows.  After integration by parts in \eqref{hypLag}, the scalar part of the action takes the form
\es{ActionQuad}{
	S =  -\f12 \int d^3x\, \sqrt{g}\, z_{iI} \mathbb{M}^{iI}{}_{jJ} z^{jJ}\,,
}
with
\es{GotM}{
	\mathbb{M}^{iI}{}_{jJ}(x) = -(\cD_\mu\cD^\mu)^I{}_J - \f14 \left( -\f12 R + D - C^2 \right) \delta^I_J +\f{1}{2}  L^p{}_q{}^I{}_K L^q{}_p{}^K{}_J +i \, Y^{i}{}_j{}^I{}_J\,,
}
where we should plug in the background values of all the fields. In terms of the differential operator $\mathbb{M}$, the Green's function $G$ can be obtained as the solution to the equation
\begin{equation}
	\mathbb{M}^{iI}{}_{jJ}(x) G^{jJ}{}_{kK}(x,x^\prime) = -\f{\delta^i_k \delta^I_K}{\sqrt{g(x^\prime)}}\delta^{(3)}(x-x^\prime)\,.
\end{equation}
Formally, one can solve this equation by first diagonalizing $\mathbb{M}^{iI}{}_{jJ}(x)$.  If $\lambda_{\vec{n}}$ are the eigenvalues and $\phi^{iI}_{\vec{n}}(x)$ are the corresponding eigenfunctions, as in
\es{evalProblem}{
	\mathbb{M}^{iI}{}_{jJ}(x)\phi^{jJ}_{\vec{n}}(x) = \lambda_{\vec{n}} \phi^{iI}_{\vec{n}}(x) \,,
}
then the Green's function is
\es{Geigenexpr}{
	\mathbb{G}^{iI}{}_{jJ}(x,x^\prime) = -\sum_{\vec{n}} \frac{\phi_{\vec{n}}^{iI}(x)\phi_{\vec{n}}^{jJ}(x^\prime)^*}{\lambda_{\vec{n}}} \,.
}
Here,  $\vec{n}$ is a multi-index labeling the eigenvalues and eigenvectors.  This sum, however, is in general difficult to evaluate for arbitrary $x$ and $x'$, but we will nevertheless be able to use this formula to evaluate the 1d theory two-point function \eqref{GDef}.

Let us proceed to solve the eigenvalue problem \eqref{evalProblem}.  The squashed sphere \eqref{3dmetric} has $\SU(2) \times \UU(1)$ isometry, so it should be possible to write the operator $\mathbb{M}^{iI}{}_{jJ}$ in terms of the $\SU(2)$ quadratic Casimir $C_2$ and the $\UU(1)$ generator $T$
\es{C2Def}{
	C_2 = {\cal L}_{v_i^l}{\cal L}_{v_i^l} = - \partial_\theta^2  - \cot\theta  \partial_\theta
	- \frac{1}{\sin^2 \theta} \left(\partial_\phi^2 + \partial_\psi^2 - 2 \cos \theta \partial_\phi \partial_\psi  \right)   \,, \qquad
	T = - 2 i \partial_\psi \,,
}
where $v_i^l$ are the Killing vectors in \eqref{Killingv}. An explicit examination of \eqref{GotM} gives
\es{MExplicit}{
	r^2\mathbb{M}^{iI}{}_{jJ} &=  \left(4 C_2 - \left(\f{1-b^2}{1+b^2} \right)^2 T^2 + \f{b^2(2+4m^2+b^2)}{(1+b^2)^2}\right) \delta^i_j\delta^I_J\\
	&{}+ 2\f{(1-b^2)}{(1+b^2)^2}T\left((\sigma_3)^i{}_j\delta^I_J+2i\,b \, m \,\delta^i_j(\sigma_3)^I{}_J\right)- \f{4i\,b\, m}{(1+b^2)^2}(\sigma_3)^i{}_j(\sigma_3)^I{}_J\,.
}

The eigenfunctions and eigenvalues of $\mathbb{M}^{iI}{}_{jJ}$ can be determined as follows.  To simplify notation, let us introduce the embedding coordinates
\es{EmbeddingCoords}{
	w_1 &= \cos\tfrac{\theta}{2}\,\e^{\f i2 (\phi+\psi)}\,, \qquad 
	w_2 = \sin\tfrac{\theta}{2}\,\e^{\f i2 (-\phi+\psi)}\,.  
}
It is straightforward to check that $\begin{pmatrix} w_1 \\ w_2 \end{pmatrix}$ and $\begin{pmatrix} -\bar w_2 \\ \bar w_1 \end{pmatrix}$  form doublets under $\SU(2)$---hence they have $\SU(2)$ spin $\ell = 1/2$ and $C_2 = \ell(\ell+1) = 3/4$---and have $T$ charges $+1$ and $-1$, respectively.  By taking tensor products of these $\SU(2)$ doublets, one can construct simultaneous eigenfunctions of $C_2$ and $T$, with eigenvalues $\ell(\ell+1)$ under $C_2$ and $T$ eigenvalues ranging from $-2\ell$ to $2\ell$ in even steps.  A complete basis of normalizable functions on the squashed sphere can thus be labeled as $\psi_{\ell m_\ell n}$, with $\ell \in \f12\mathbb{Z}_{\geq 0}$ and $m_\ell = -\ell, -\ell + 1, \cdots, \ell$ labeling the $\SU(2)$ quantum numbers, and $n = -2\ell, -2\ell + 2, \cdots, 2\ell$ being the charge under $T$.  There is a unique such function for every $\ell, m_\ell, n$.   The eigenvalues and eigenfunctions of $\mathbb{M}^{iI}{}_{jJ}$ are then labeled by the quantum numbers $(\ell, m_\ell, n)$, as before, and also by the eigenvalues $\pm$ under $(\sigma_3)^i{}_j$ and $(\sigma_3)^I{}_J$.  The eigenvalues of $\mathbb{M}^{iI}{}_{jJ}$ are easy to read off  from \eqref{MExplicit} by replacing $C_2 \to \ell(\ell+1)$ and $T \to n$, as well as $(\sigma_3)^i{}_j \to s_1 \in \{\pm\}$ and $(\sigma_3)^I{}_J \to s_2 \in \{ \pm \}$:\footnote{The multi-index $\vec{n}$ appearing in \eqref{evalProblem}--\eqref{Geigenexpr} is $\vec{n} = (\ell, m_\ell, n, s_1, s_2)$.}
\es{lambda}{
	\lambda_{\ell, m_\ell, n}^{s_1 s_2} = \frac{4}{r^2} \left[\ell + \frac{b^2 - 2 i b s_1 s_2 m + (1 - b^2) s_1 n }{2(b^2 + 1)} \right] 
	\left[\ell + 1 - \frac{b^2 - 2 i b s_1 s_2 m + (1 - b^2) s_1 n }{2(b^2 + 1)} \right] \,,
}
with the corresponding eigenfunctions being 
\es{Efns}{
	(\phi_{\ell, m_\ell, n}^{++})^{iI} &= \psi_{\ell m_\ell n} \delta^i_1 \delta^I_1 \,, \qquad
	(\phi_{\ell, m_\ell, n}^{+-})^{iI} = \psi_{\ell m_\ell n} \delta^i_1 \delta^I_2 \,, \\
	(\phi_{\ell, m_\ell, n}^{-+})^{iI} &= \psi_{\ell m_\ell n} \delta^i_2 \delta^I_1 \,, \qquad
	(\phi_{\ell, m_\ell, n}^{--})^{iI} = \psi_{\ell m_\ell n} \delta^i_2 \delta^I_2 \,.
} 

We can then combine \eqref{GDef} and \eqref{Geigenexpr} to compute the 1d two-point function 
\es{GDef2}{
	G^{I}{}_J(\alpha_{12})  = u_i(\alpha_1) u^j(\alpha_2)\sum_{\vec{n}} \frac{\phi_{\vec{n}}^{iI}(\alpha_1)\phi_{\vec{n}}^{jJ}(\alpha_2)^*}{\lambda_{\vec{n}}} \,,
}
where the $\phi^{iI}$ are evaluated at $\theta = 0$ and angular coordinates $\alpha_i$.  Note that when $\theta = 0$, we have $w_2 = 0$ and $w_1 = e^{i \alpha}$, so the only eigenfunctions that contribute to the sum \eqref{GDef} are those that do not vanish when $w_2 = 0$.  As mentioned above, one can construct all eigenfunctions as polynomials in $w_i$ and $\bar w_i$, and one can show that the normalized eigenfunctions that do not vanish when $w_2 = 0$ are
\begin{equation}
	f_{p,q} = \f{1+p+q}{r^3\pi^2 (b+b^{-1})} \sum_{k=0}^q (-1)^k \binom{p}{k}\binom{q}{k} w_1^{p-k}\bar{w}_1^{q-k}w_2^{k}\bar{w}_2^{k}\,,
\end{equation}
with $\ell = (p + q)/2$ and $n = p - q$.

Performing the sums in \eqref{GDef2} is rather onerous, but the resulting Green's function is given by\footnote{At coincident points, we take $G^I{}_J (0) = i \f{\tanh \f{\pi m}{b}}{8\pi r b} \delta^I_J$.} 
\es{1dGreen}{
	G^{I}{}_{J}(\alpha_{12})\equiv\langle \cZ^I (\alpha_1) \cZ_J (\alpha_2) \rangle = \left( i\f{\, \sgn \alpha_{12} {\bf 1}_2 -\tanh\f{\pi m \sigma_3}{b}}{8\pi b r}\e^{\f{m \sigma_3}{b}\alpha_{12}} \right)^I_{\phantom{I}J}  \,,
}
where ${\bf 1}_2$ is the $2 \times 2$ identity matrix. Note that the final answer only depends on $b$ in a very simple way and is therefore closely related to the round sphere result \cite{Dedushenko:2016jxl}.

While this formula was derived for a single hypermultiplet, it is straightforward to generalize it to $n_h$ free hypermultiplets.  This theory has an $\USp(2n_h)$ flavor symmetry, which can be coupled to a background vector multiplet that can be given the supersymmetry-preserving values in   \eqref{VectorBk} with  $\mu = m_a T_F^a$, where $T_F^a$ are the $\USp(2 n_h)$ generators.  In this case, the two point function is 
\es{1dGreenFreeHypers}{
	G^{I}{}_{J}(\alpha_{12})\equiv\langle \cZ^I (\alpha_1) \cZ_J (\alpha_2) \rangle = \left( i\f{\, \sgn \alpha_{12} {\bf 1}_{2n_h} -\tanh\f{\pi m_a T_F^a}{b}}{8\pi b r}\e^{\f{m_a T^a}{b}\alpha_{12}} \right)^I_{\phantom{I}J} \,,
}
where now ${\bf 1}_{2n_h}$ is the $2n_h \times 2 n_h$ identity matrix.  To derive \eqref{1dGreenFreeHypers}, we can first perform an $\USp(2n_h)$ transformation to put $m_a T_F^a$ in block diagonal form, with each block proportional to $\sigma_3$.  Then we can use the result \eqref{1dGreen} for each hypermutiplet.  

Before moving on to discussing gauge theories, let us ask whether it is possible to write down a 1d theory that reproduces the two-point function \eqref{1dGreenFreeHypers}.  Since the two-point function \eqref{1dGreenFreeHypers} is the same as that in \cite{Dedushenko:2016jxl} with extra factors of $b$ sprinkled around, it is not hard to see that the 1d theory that reproduces \eqref{1dGreenFreeHypers} has partition function
\es{1dFreeHyperTheory}{
	Z_\text{1d} = \int \prod_{I=1}^{2n_h} \cD \cZ^I\,\,  \exp\left\{ 4\pi i r \, \int \dd\alpha  \left( b \cZ_I\partial_\alpha \cZ^I + \cZ_I m_a  (T^a)^I{}_J \cZ^J \right)\right\} \,.
}
To show this, note that the Green's function \eqref{1dGreenFreeHypers} obeys
\es{GreenEq}{
	-8 \pi i  r ( b {\bf 1}_{2n_h}  \partial_\alpha - m_a T^a)^I{}_J G^J{}_K (\alpha) = \delta^I_K  \delta(\alpha) \,,
}
which is the equation that follows from \eqref{1dFreeHyperTheory}. An alternative path to reaching this 1d action was described in \cite{Panerai:2020boq} by explicitly using equivariant localization of the 3d action.

\subsection{Gauge theories}

Next, let us consider the 1d sector of gauge theories with dynamical vector multiplets and hypermultiplets.  The dynamical vector multiplets gauge a subgroup $G$ of the $\USp(2 n_h)$ symmetry of $n_h$ free hypermultiplets.  Within $\USp(2n_h)$, the commutant of $G$ is the flavor symmetry group $G_F$, for which one can turn on real mass parameters $m_a T^a_F$ as above.  In this case, the action is simply 
\es{ActionGauge}{
	S[\cV,\cH]= S_{\rm hyp}[\cH^I, \cV + \cV_\text{bk} \big|_{\mu = m_a T_F^a}] + S_{\rm YM}[\cV]\,.
}

Using the supersymmetry variations (see Appendix~\ref{app:confSugra}), one can show that the Yang-Mills action $S_\text{YM}$ can be written as
\begin{equation}
	S_{\rm YM} = \delta_{+}\delta_{-}\left( \f{i}{2g_{\rm YM}^2}(\sigma_3)^j{}_i (\sigma_3)^q{}_p \int\dd^3 x \,\sqrt{g}\Tr\left\{ \Omega^{ip}\Omega_{jq} - 2 L^p{}_q Y^i{}_j \right\}\right)\,,
\end{equation}
where $\delta_+$ is the supersymmetry transformation associated to $\cQ^H$ and $\delta_-$ the transformation associated to the supercharge obtained from $\cQ^H$ by flipping the relative sign between the two terms. From this relation it immediately follows that the Yang-Mills action is $\cQ^H$-exact.  An immediate consequence of this fact is that correlation functions of the 1d operators will be independent of the Yang-Mills coupling.

Owing to the fact that the Yang-Mills action is positive-definite, we can take $g_\text{YM} \to 0$, and in this limit the dynamical vector multiplet localizes on configuration on which $S_{\rm YM}$ vanishes.  These configurations are nothing but the supersymmetric backgrounds \eqref{VectorBk} we found in Section~\ref{subsec:NONCONFORMAL}, namely
\es{VRestriction}{
	\cV = \cV_\text{bk} \big|_{\mu = \sigma} \,,
}
where $\sigma$ lies within the Lie algebra of $G$\@. The entire contribution from the vector multiplet comes from the one-loop determinant of fluctuations of the vector multiplet field around the configuration \eqref{VRestriction}, which we now turn to.

\subsection{One-loop determinants} 
\label{DETERMINANTS}

In this subsection, we determine the one-loop determinant of fluctuations of a dynamical vector multiplet with action $S_{\rm YM}[{\cal V}]$ expanded around the supersymmetric configuration \eqref{VRestriction}, as well as the one-loop determinant of fluctuation of a hypermultiplet coupled to a background vector multiplet which is given the supersymmetric profile $\cV_\text{bk} \big|_{\mu = \sigma}$.  The first computation is relevant for the localization of the dynamical vector multiplet, while the latter will be useful for determining an action for the 1d sector of general gauge theories.  We will start with the hypermultiplet because we have already computed the eigenvalues of the scalar fluctuations in \eqref{lambda}.

Given that the ${\cal N} = 4$ gauge theories we study here are particular cases of ${\cal N} = 2$ theories, and that general ${\cal N} = 2$ theories on the squashed sphere were studied in \cite{Hama:2011ea,Imamura:2011wg}, we can determine the one-loop determinants from appropriately combining these ${\cal N} = 2$ results.  On general grounds, the ${\cal N} = 2$ R-symmetry is the diagonal linear combination of the $\UU(1)_H$ and $\UU(1)_C$ Cartans of $\SU(2)_H$ and $\SU(2)_C$, respectively, normalized such that the R-charges of the highest weight states of $\SU(2)_H$ or $\SU(2)_C$ fundamentals are equal to $1/2$.  The difference of the two Cartans is a flavor symmetry $F$ from the ${\cal N} =2$ point of view. Thus, we have the following decompositions:
\begin{equation}\label{Decompositions}
	\begin{array}{lcl}
		\hspace{0mm}\text{\underline{${\cal N} = 4$ multiplets}} & & \hspace{25mm}\text{\underline{${\cal N} = 2$ multiplets}} \\[5mm]
		{\rm Hypers}:\,   & \longrightarrow & \hspace{5.1mm}\left[{\rm Chirals:}\, (z^{1I}, \zeta^{I2})\,\, {\rm w/}\,\, (R, F) = (1/2, 1)\right] \\
		(z^{iI}, \zeta^{Ip}) & &{}\oplus \left[{\rm Antichirals:}\, (z^{2I}, \zeta^{I1})\,\, {\rm w/}\,\, (R, F) = (-1/2, -1)\right] \\[5mm]
		{\rm Twisted \,\, hypers:}\,  & \longrightarrow & 
		\hspace{5.1mm}\left[{\rm Chirals:}\, (\wti z^{1 \tilde I}, \wti \zeta^{\tilde I2})\,\, {\rm w/} \,\,(R, F) = (1/2, -1)\right] \\[1mm]
		 (\wti z^{p \tilde I}, \wti \zeta^{\tilde Ii}) & &{}\oplus  \left[{\rm Antichirals:} \,(\wti z^{2 \tilde I}, \wti \zeta^{\tilde I1}) \,\, {\rm w/}\,\, (R, F) = (-1/2, 1)\right] \\[5mm]
		{\rm Vector:} \,  & \longrightarrow & \hspace{5.1mm}\left[{\rm Vector:}\, (L^{12}, Y^{12}, A_\mu, \Omega^{11}, \Omega^{22})\,\, {\rm w/}\,\, F=0 \right]  \\
		(L^{pq}, Y^{ij}, A_\mu, \Omega^{ip}) & &{}\oplus 
		\left[{\rm Chiral:}\, (L^{11}, \Omega^{21}, Y^{22}) \,\,{\rm w/}\,\, (R, F) = (1, -2)\right] \\
		& &{}\oplus  \left[{\rm Antichiral:}\, (L^{22}, \Omega^{12}, Y^{11})\,\, {\rm w/}\,\, (R, F) = (-1, 2)\right]  \\[5mm]
		{\rm Twisted\,\, vector:} \,    
		& \longrightarrow & \hspace{5.1mm}\left[{\rm Vector:} \,(\wti L^{12}, \wti Y^{12}, \wti A_\mu, \wti \Omega^{11}, \wti \Omega^{22})\,\, {\rm w/} \,\, F=0 \right] \\[1mm]
		(\wti L^{ij}, \wti Y^{pq}, \wti A_\mu, \wti \Omega^{ip}) & &{}\oplus 
		\left[{\rm Chiral:}\, (\wti L^{11}, \wti\Omega^{21}, \wti Y^{22}) \,\,{\rm w/} \,\, (R, F) = (1, 2)\right]\\[1mm]
		& &{}\oplus  \left[{\rm Antichiral:}\, (\wti L^{22}, \wti \Omega^{12}, \wti Y^{11})\,\, {\rm w/}\,\, (R, F) = (-1, -2)\right]		
	\end{array}
\end{equation}

A potential challenge is that there are two $\SU(2) \times \UU(1)$-preserving ${\cal N} = 2$ backgrounds:  the one in \cite{Hama:2011ea} has the property that the partition function is independent of the squashing parameter $b$, while for the background in \cite{Imamura:2011wg} the partition function depends non-trivially on $b$.  Our ${\cal N} = 4$ background corresponds to the latter, as can be seen from matching the form of the hypermultiplet scalar eigenvalues \eqref{lambda} to the eigenvalues of the scalars in the ${\cal N} = 2$ chiral multiplets given in \cite{Imamura:2011wg}.\footnote{It is also straightforward to check that the eigenvalues in \eqref{lambda} do not match the eigenvalues in \cite{Hama:2011ea} in that the relative coefficient of $\ell(\ell+1)$ and $n^2$ is different in that case.}

The one-loop determinants for ${\cal N} = 2$ chiral\footnote{The determinant corresponding to a chiral multiplet includes the contribution from its conjugate anti-chiral multiplet.} and vector multiplets were computed in \cite{Imamura:2011wg}.  Let us consider a chiral multiplet transforming in representation $\cR' \otimes \cR_F'$ of a product $G' \times G_F'$ of gauge and flavor symmetries,\footnote{We use primes here to denote properties of the ${\cal N} = 2$ theory.} coupled to both dynamical  and background vector multiplets. The dynamical and backgrounds vector multiplets are restricted to supersymmetric configurations parameterized by constant values $\sigma'$ of a scalar in the vector multiplet and real mass parameters $m'$.  Here, $\sigma'$ and $m'$ belong to the Cartans of the Lie algebras $\fg'$ of $G'$ and $\fg_F'$   of $G_F'$, respectively.  For a chiral multiplet of R-charge $R$, the one-loop determinant is \cite{Imamura:2011wg}
\es{DeltaChiral}{
	\Delta_\text{chiral}^{\cR', \cR_F'} (R, \sigma', m') = \prod_{(\rho', \rho_F') \in (W_{\cR'}, W_{\cR_F'}) } \frac{1}{ s_b \left( \frac{(b + b^{-1})\left( \rho' \cdot \sigma'  + \rho_F' \cdot m' - i (1 - R) \right) }{2}  \right) }
}
where $W_{\cR}$ denotes the set of weights of the representation $\cR$.  Here, $s_b$ is the double sine function defined as
\es{DoubleSine}{
	s_b(x) \equiv \prod_{m, n \geq 0} \frac{ m b + n b^{-1} + \frac{b + b^{-1}}{2} - ix}{m b + n b^{-1} + \frac{b + b^{-1}}{2} + ix} \,.
}
For an ${\cal N} = 2$ vector multiplet, one can write a similar formula in terms of a product over the roots of $\fg'$.  Denoting by $W_\text{adj}'$ the set of roots, we have
\es{DeltaN2Vector}{
	\Delta_\text{${\cal N} = 2$ vector}(\sigma') = \prod_{\rho' \in W_\text{adj}'} s_b \left( \frac{(b + b^{-1}) (\rho' \cdot \sigma' - i) }{2}  \right)  \,.
}

For a hypermultiplet that decomposes as a chiral in representation $(\cR \otimes \cR_F) \oplus (\overline{\cR} \otimes \overline{\cR}_F)$ of $G \times G_F$ and which has flavor charge $F$ under the $\UU(1)$ flavor symmetry mentioned above, we have
\es{rhop}{
	\rho' &= \rho \qquad \, \ \ \rho_F' = (F, \rho_F) \,, \hspace{12.5mm}\text{for chiral in $\cR \otimes \cR_F$} \,, \\
	\rho' &= -\rho \qquad \rho_F' = (F, -\rho_F) \hspace{12mm} \text{for chiral in $\overline{\cR}  \otimes \overline{\cR}_F$}  \,.
} 
By matching the scalar eigenvalues we identify
\es{sigmap}{
	\sigma' = \frac{2\sigma}{b + b^{-1}} \,, \qquad
	 m' = \left( \frac{2 m_F}{b + b^{-1}}  ,\frac{2 m}{b + b^{-1}} \right) \,, \qquad m_F \equiv \frac{i (b - b^{-1}) }{4} \,,
}
so that we have
\es{Prod}{
	\rho' \cdot \sigma' = \frac{2\rho \cdot \sigma}{b + b^{-1}}  \,, \qquad
	 \rho' \cdot m' =  \frac{2\rho \cdot m}{b + b^{-1}}  + \frac{i}{2} \frac{b - b^{-1}}{b + b^{-1}} f \,.
} 
From the ${\cal N} = 2$ point of view, the term proportional to the charge $F$ corresponds to a real mass $m_F$ for this flavor symmetry that is needed in order to preserve ${\cal N}= 4$ supersymmetry  (See also \cite{Minahan:2021pfv}).  We can plug \eqref{Prod} into \eqref{DeltaChiral} with appropriate values for $R$ and $F$.   

A hypermultiplet consists of a chiral multiplet with $(R, F) = (\frac 12, 1)$ and an antichiral multiplet with the same charges $(R, F)$ but $\rho \to -\rho$. The resulting one-loop determinant is given by:
\es{DeltaHyper}{
	\Delta_\text{hyper}(\sigma) &=  \prod_{(\rho, \rho_F) \in (W_\cR, W_{\cR_F})} \frac{1}{ s_b \left( \rho \cdot \sigma  + \rho_F \cdot m - \frac{i}{2b}   \right) s_b \left( -\rho \cdot \sigma -  \rho_F \cdot m - \frac{i}{2b}   \right) } \\
	 &= \prod_{(\rho, \rho_F) \in (W_\cR, W_{\cR_F})}   \frac{1}{2 \cosh( \frac{\pi \,\rho \cdot \sigma + \pi\, \rho_F \cdot m}{b})} \,,
}
where the second equality can be derived from the definition \eqref{DoubleSine} of the double sine function. 
Similarly, for a twisted hypermultiplet we have a factor with $(R, F) = (\frac 12, -1)$ and another factor with the same values of $(R, F)$ but $\rho \to -\rho$:
\es{DeltaTwistedHyper}{
	\Delta_\text{twisted hyper}(\sigma) &=  \prod_{(\rho, \rho_F) \in (W_\cR, W_{\cR_F})} \frac{1}{ s_b \left( \rho \cdot \sigma  + \rho_F \cdot m - \frac{ib}{2}   \right) s_b \left( -\rho \cdot \sigma -  \rho_F \cdot m - \frac{ib}{2}   \right) } \\
	 &= \prod_{(\rho, \rho_F) \in (W_\cR, W_{\cR_F})}   \frac{1}{2 \cosh( \pi b \,\rho \cdot \sigma + \pi b\, \rho_F \cdot m)} \,,
}

An ${\cal N} = 4$ vector multiplet is a direct sum of an ${\cal N} = 2$ vector multiplet and an adjoint chiral multiplet with $(R, F) = (1, -2)$.  For an ${\cal N} = 2$ vector with $F = 0$, from \eqref{DeltaN2Vector} and \eqref{sigmap} we obtain
\begin{equation}\label{DeltaN2VectorAgain}
	\begin{aligned}
		\Delta_\text{${\cal N} = 2$ vector}(\sigma) &= \prod_{\alpha \in W_\text{adj}'} s_b \left(   \sigma-i \frac{(b + b^{-1}) }{2}  \right)\\
		&= \prod_{\alpha \in W_\text{adj}'} \sqrt{ 4 \sinh (\pi b \alpha \cdot \sigma) \sinh \left( \frac 1b \pi \alpha \cdot \sigma \right)} \,.
	\end{aligned}
\end{equation}
For an ${\cal N} = 2$ chiral with $(R, F) = (1, -2)$, we have 
\es{adjChiral}{
	\Delta_{\text{adj chiral}}(\sigma) = \prod_{\alpha \in W_\text{adj}} \frac{1}{s_b \left( \alpha \cdot \sigma - \frac{i(b - b^{-1} )}{2}  \right) }
	= \frac{1}{b^{|G|}} \prod_{\alpha \in W_\text{adj}'} \sqrt{\frac{\sinh \left(\frac 1b \pi \alpha \cdot \sigma \right) }{\sinh (\pi b \alpha \cdot \sigma)}} \,,
}
where $|G|$ is the dimension of the Cartan subalgebra of $G$.  Multiplying \eqref{DeltaN2VectorAgain} and \eqref{adjChiral} we obtain the contribution from an ${\cal N} = 4$ vector multiplet:
\es{DeltaN4Vector}{
	\Delta_{\text{${\cal N} = 4$ vector}}(\sigma) = \frac{1}{b^{|G|}} \prod_{\alpha \in W_\text{adj}'} 2 \sinh \left( \frac 1b \pi \alpha \cdot \sigma \right) \,.
}
For a twisted vector multiplet, we need to replace the adjoint chiral of $(R, F) = (1, -2)$ with an adjoint chiral with $(R, F) = (1, 2)$.  This simply gives the reciprocal of \eqref{adjChiral}, so in the end we find
\es{DeltaN4TVector}{
	\Delta_{\text{${\cal N} = 4$ twisted vector}}(\sigma) = b^{|G|} \prod_{\alpha \in W_\text{adj}'} 2  \sinh \left(  \pi b\,\alpha \cdot \sigma \right) \,.
}

\subsection{A 1d theory for the Higgs branch}

We can now put all these ingredients together in order to write down a succinct description of the 1d Higgs branch theory of 3d gauge theories with vector multiplets and hypermultiplets. In the general case, the squashed sphere partition function of these theories is
\es{ZS3Combined}{
	Z_{S^3_b} =  \frac{1}{|\cW|} \int_\fh d\sigma_a \,  \Delta_b(\sigma)
	Z_\text{1d} \,, \qquad  \Delta_b(\sigma) \equiv \frac{1}{b^{|G|}} \prod_{\alpha \in W_\text{adj}'} 2  \sinh \left(  \frac{1}{b} \pi \alpha \cdot \sigma \right) 
}
where the integration is over the Cartan $\fh$ of the Lie algebra $\fg = \text{Lie}(G)$,  $|\cW|$ is the order of the Weyl group of $\fg$, and, as above, $Z_\text{1d}$ is given by the contribution of the hypermultiplet
\es{GotZ1d}{
	Z_\text{1d}(\sigma, m) = \prod_{(\rho, \rho_F) \in (W_\cR, W_{\cR_F})}   \frac{1}{2 \cosh\left( \frac{\pi }{b}  (\rho \cdot \sigma +  \rho_F \cdot m) \right) }  \,.
}
However, $Z_{1d}$ can be written as the one-dimensional Gaussian theory in \eqref{1dFreeHyperTheory}, namely
\es{1dGaugedHypers}{
	Z_\text{1d}(\sigma, m) = \int \prod_{I=1}^{2n_h} \cD \cZ^I\,\,  \exp\left\{ 4\pi i r \, \int \dd\alpha  \left( b \cZ_I\partial_\alpha \cZ^I + \cZ_I (\sigma_a T^a + m_a  T_F^a)^I{}_J \cZ^J) \right)\right\} \,,
}
provided that the integration contour of the path integral is over a middle-dimensional integration cycle, as in the round sphere case discussed in \cite{Dedushenko:2016jxl}.  In line with the previous discussions, for every $\UU(1)$ factor in the gauge group, we can introduce an FI term, which leads to an additional insertion of $\e^{-8\pi^2\, i\, b\,\zeta\,\sigma}$ in \eqref{ZS3Combined}. 

The advantage of writing the partition function as a one-dimensional Gaussian theory coupled to a matrix model is that in this formulation we get access to a much wider range of observables. Indeed, the one-dimensional theory can now be used to calculate correlation functions of the twisted Higgs branch operators inserted along the circle parameterized by $\alpha$ as follows,
 \es{CorrelationFunctions}{
	\left\langle  \cO_1(\alpha_1)\cdots \cO_n(\alpha_n) \right\rangle_{b, m} = \f{1}{Z_{S^3_b}|\cW|}\int_{\fh} \dd\sigma\, \Delta_b(\sigma) \,\left\langle  \cO_1(\alpha_1)\cdots \cO_n(\alpha_n) \right\rangle_\sigma\, Z_\text{1d}(\sigma, m)\,,
}
where $\langle\cdots\rangle_\sigma$ denotes the correlation function computed in the 1d theory. Since the latter is Gaussian we can simply compute these correlation functions using Wick contractions with the propagator $G^I{}_J(\alpha_1-\alpha_2)$ computed in \eqref{1dGreenFreeHypers}.

Note that the dependence on $b$ of the correlation functions can be removed after field redefinitions.  In particular, after redefining $m \to b \,m$ and $\cZ \to \frac{1}{\sqrt{b}} \cZ$, then after the change of variables $\sigma \to b \,\sigma$ in \eqref{ZS3Combined} and \eqref{CorrelationFunctions} the entire $b$ dependence drops out.  Thus, if we did not perform these redefinitions and change of variables, we would conclude that the $b$ dependence can be inferred from the scaling dimensions $\Delta_i$ of the 3d operators from which the $\cO_i$ originate:
 \es{bScaling}{
 \left\langle  \cO_1(\alpha_1)\cdots \cO_n(\alpha_n) \right\rangle_{b, m}
  =  b^{\sum_{i=1}^n \Delta_i} \left( \left\langle  \cO_1(\alpha_1)\cdots \cO_n(\alpha_n) \right\rangle_{1, m} \bigg|_{m \to m/b} \right) \,.
 }

As was shown in \cite{Dedushenko:2016jxl}, it is possible to interpret this one-dimensional Gaussian theory, coupled to a matrix model as a gauge-fixed gauged quantum mechanics. This can be seen by rewriting the partition function \eqref{ZS3Combined} as follows. First, we rewrite the integral as an integral over the full Lie algebra instead of the Cartan. Doing so we introduce an additional Vandermonde determinant and remove the factor $|\cW|$. Noting then that,
\begin{equation}
	\f{2\sinh \left( \frac{\pi}{b} \alpha\cdot\sigma \right)}{\alpha\cdot\sigma} = \cC \prod_{n=1}^{\infty} \left(n^2+ \frac{\alpha\cdot\sigma}{b} \right) \,,
\end{equation}
where $\cC$ is a divergent $\alpha\cdot\sigma$-independent normalization factor. It then immediately follows, analogous to \cite{Dedushenko:2016jxl}, that we can rewrite \eqref{ZS3Combined} as
\begin{equation}
	Z_{S^3_b} = \int \frac{\dd\sigma}{b}\,  Z_\text{1d}(\sigma)\int D^\prime c\,  D^\prime \wti c \,\exp\left[ - \int \dd\alpha \,\wti c\, \partial_\alpha \left(\partial_\alpha + \frac{\alpha\cdot\sigma}{b}  \right)\,c \right]\,.
\end{equation}
However, this additional factor can simply be interpreted as the Faddeev-Popov ghost action \cite{Faddeev:1967fc} corresponding to a gauge fixing condition $\partial_\alpha\cA_\alpha = 0$ for a 1d gauge field $\cA$, solved by $\cA_\alpha =\sigma/b$. Combining the above, we find that when the real masses and FI parameters vanish, our theory is described by a gauge-fixed version of the following gauged quantum mechanics
\begin{equation}
	Z_{S^3_b} = \int \cD \cA\, \cD \cZ^I \exp\left[ 4\pi i r\,b \int \dd\alpha\,  \cZ_I \cD_\alpha \cZ^I \right]\,,\qquad \qquad \cD_\alpha \equiv \partial_\alpha +  \cA_\alpha\,.
\end{equation}
In particular, what this formulation teaches us is that
\begin{equation}\label{Dterm}
	\left( \cZ_I \cR(T)\cZ^I \right)(\alpha) = 0 \,,\qquad\qquad \text{for all } T\in \fg\,
\end{equation}
up to contact terms.  When non-trivial real masses and FI parameters are turned on, the covariant derivative gets modified to $\cD_\alpha = \partial_\alpha + \cA_\alpha + \f{mr}{b}$ and the path integral acquires an extra factor $\exp\left[ -8\pi^2\,i\,b\,{\rm Tr}_{\zeta}\cA \right]$. In addition, in the presence of non-vanishing FI parameters, the right hand side of \eqref{Dterm} receives a contribution proportional to the FI parameters.

\subsection{Theories with twisted vector and hypermultiplets}

One can now go through the same steps to derive the 1d theory for a gauge theory with twisted vector multiplets and twisted hypermultiplets. In this case the partition function takes the form
 \es{TwistedTheory}{
	\wti Z_{S^3_b} = \frac{1}{|\wti \cW|} \int_\fh d \wti \sigma_a \, \wti \Delta_b(\wti \sigma) 
	\wti Z_\text{1d} \,, \qquad  \wti \Delta_b(\wti \sigma) \equiv b^{|\wti G|} \prod_{\wti \alpha \in W_{\wti{ \text{adj}}}'} 2  \sinh \left(  \pi b\wti \alpha \cdot \wti \sigma \right) \,,
}
where $|\wti \cW|$ is the order of the Weyl group of the gauge algebra $\wti \fg = \text{Lie}(\wti G)$ associated to the dynamical twisted vector multiplets.The 1d partition function of a twisted hypermultiplet is
\es{GotZ1dTwisted}{
	\wti Z_\text{1d} = \prod_{(\wti \rho, \wti \rho_F) \in ( W_{\wti \cR},W_{\wti \cR_F})}   \frac{1}{2 \cosh( \pi b \, \wti \rho \cdot \wti \sigma + \pi b\, \wti \rho_F \cdot \wti m)}  \,.
}
where $\wti m$ are the real mass parameters.  The resulting theory \eqref{TwistedTheory} differs from \eqref{ZS3Combined} only in the replacement $b \to 1/b$, and thus everything we mentioned above also holds in this case, provided one makes such a replacement.  In particular, the 1d partition function appearing in \eqref{TwistedTheory} can be written as
 \es{1dGaugedTwistedHypers}{
	\wti Z_\text{1d} = \int \prod_{I=1}^{2n_h} \cD \wti \cZ^I\,\,  \exp\left\{ 4\pi i r \, \int \dd\alpha  \left( \frac{1}{b} \wti \cZ_I\partial_\alpha \wti \cZ^I + \wti \cZ_I (\wti \sigma_a \wti T^a + \wti m_a  \wti T_F^a)^I{}_J \wti \cZ^J) \right)\right\} \,,
}
and the $b$ dependence is as in \eqref{bScaling}, with $b \to 1/b$.

%%%%%%%%%%%%%%%%%%%%%%%%%%%%%%%%%%
\section{Mass deformations of the 1d theory}
\label{sec:genericMassDef}
%%%%%%%%%%%%%%%%%%%%%%%%%%%%%%%%%%

So far, we described a cohomological construction for the one-dimensional theories, $\cT_H$ and $\cT_C$, in Section~\ref{sec:1dSectorCohomology}, and then, in Section~\ref{sec:1dSector}, we discussed in detail the resulting $\cT_H$ theory arising in 3d theories containing either vector and hypermultiplets or twisted vector and twisted hypermultiplets. However, the cohomological construction in Section~\ref{sec:1dSectorCohomology} applies beyond this class of examples. In this section, we return to this more general setup and comment on some additional properties of the 1d theories present for any 3d $\cN=4$ QFT in the $\su(2|1)\times \psu(1|1)$ squashed sphere background, irrespective of their Lagrangian descriptions.  

In particular, we show that real mass deformations in the 3d theory are equivalent, up to $\cQ^H$-exact terms, to analogous deformations in the 1d $\cT_H$ sector.  Abstractly, a real mass deformation is obtained by coupling a conserved current multiplet $\cJ$ to a background vector multiplet $\cV_\text{bk}$ which is given the supersymmetry-preserving expectation values in \eqref{VectorBk}, with $\mu = m$:
 \es{RealMassGen}{
	S_{m}^{\text{3d}} = 
	m \int d^3x\, \sqrt{g}\, \left[ \f{b^{-1}-b}{b^{-1}+b}j^3   - \frac{1}{2br} (\sigma_3)^i{}_j J^j{}_i + \frac{i}{2} \left(\sigma_3\right)^{p}{}_q K^q{}_p  \right] \,.
 }
We will show that 
 \es{Equivalence}{
  S_m^\text{3d} = S_m^\text{1d} + \text{(${\cal Q}_H$-exact term)} \,,
 }
where
\begin{equation}\label{RealMass1d}
	S_{m}^{\text{1d}} = \pi  m r^2 \int_{-\pi}^\pi d \alpha \left[ J^1{}_1 - e^{i \alpha} J^1{}_2 + e^{-i \alpha} J^2{}_1 - J^2{}_2 \right] \,.
\end{equation}
What this means is that insertions of $S_{m}^{\text{3d}}$ into correlation functions of supersymmetric operators are equivalent to insertions of $S_m^\text{1d}$.\footnote{ An analogous equivalence holds when a theory has a twisted flavor symmetry $\wti G_F$ is deformed by a twisted  real mass parameter $\wti m$ by coupling the twisted conserved flavor current multiplet $\wti \cJ$ to a background twisted vector multiplet $\wti \cV$. }  One can argue for the validity of \eqref{Equivalence} in theories with vector multiplets and hypermultiplets using the explicit description of the 1d theory, but not in the more general class of ${\cal N}= 4$ theories that also include non-Abelian Chern-Simons interactions.  Eq.~\eqref{Equivalence} was used in \cite{Agmon:2017lga,Agmon:2017xes,Agmon:2019imm,Binder:2019mpb,Binder:2020ckj} and checked in a perturbative weak-coupling expansion in ABJM theory in \cite{Gorini:2020new}.  It is analogous to the equivalence between deformations of the $S^4$ partition functions by the integrated top component of a chiral multiplet and the insertion of the bottom component of the chiral multiplet at the North and South poles of the sphere \cite{Gomis:2014woa,Gerchkovitz:2016gxx}.  In the supersymmetric background of Section~\ref{FIRSTBACKGROUND}, the SUSY parameter $\epsilon$ corresponding to the $\cQ^H$ supercharge is
 \es{epsQH}{
  \epsilon^{11} &= \left( \sqrt{b}\,, 0 \right)\,, \qquad\hspace{2mm}
  \epsilon^{12} = \left( \f{\e^{\f i2 (\phi-\psi)}\sin\f{\theta}{2}}{\sqrt{b}}\,, \sqrt{b}\e^{\f i2 (\phi+\psi)}\cos\f{\theta}{2} \right)\,, \\
   \epsilon^{22} &=  \left( 0 \,, -\sqrt{b} \right)\,. \qquad \epsilon^{21} = \left( -\sqrt{b}\e^{-\f i2 (\phi+\psi)}\cos\f{\theta}{2} \,, \f{\e^{-\f i2 (\phi-\psi)}\sin\f{\theta}{2}}{\sqrt{b}} \right)\,.
 }
What we would like to show is that the integrand of \eqref{RealMassGen} is a $\cQ^H$-exact term, namely $\bar \epsilon_{ip}' \delta \Xi^{ip}$ for some coefficients $\epsilon_{ip}'$, plus a total derivative everywhere away from the circle at $\theta = 0$.  It is important that such a relation should fail precisely at $\theta = 0$, because otherwise $S_m^\text{3d}$ would be $\cQ^H$-exact and therefore its insertion in supersymmetric correlators would produce a vanishing result.

Using the supersymmetry variations of the conserved current multiplet, given in \eqref{currentSUSY} one can show that it is possible to write
 \es{QExactPre}{
  S_m^{3d} = \int d^3x \, \sqrt{g} \, \left[ \bar \epsilon_{ip}' \delta \Xi^{ip} - D_\mu ( \bar \epsilon_{ip}' \gamma^\mu \epsilon^{jp} J^i{}_j ) \right]  \,, 
 }
provided that the following three conditions are obeyed,
 \es{Conditions}{
  -  i \bar \epsilon'_{ip} \gamma^c  \epsilon^{ip} -  A_\text{bk}^c \big|_{\mu = m} &=  0 \,, \qquad 
	\bar \epsilon'_{iq} \epsilon^{ip}  - \frac{i}{2} L_{\text{bk}}{}^{p}{}_q \big|_{\mu = m}= 0 \,, \\
	 - D_\mu \left(  \bar \epsilon'_{ip}  \gamma^\mu \epsilon^{jp} \right)
        - C  \bar \epsilon'_{ip}\epsilon^{jp}  +  \bar \epsilon'_{ip}\eta^{jp} -  \frac{i}{2}  Y^{\text{bk}\, j}{}_i  &= 0 \,.
 }
A bit of algebra shows that all these conditions are satisfied if one chooses 
 \es{Gotepsilonp}{
  \bar \epsilon'_{11} &= \left( 0 , - \frac{ 2 i \sqrt{b} e^{- i \psi} m \csc \theta}{b^2 + 1} \right)  \,, \qquad \ \,
  \bar \epsilon'_{12} = \left( 0 , - \frac{ i b^{3/2} e^{- i (\phi + \psi)/2} m \sec \theta}{b^2 + 1} \right) \,, \\
  \bar \epsilon'_{21} &= \left( 0 , \frac{ i \sqrt{b}  e^{ i (\phi - \psi)/2} m \csc \theta}{b^2 + 1} \right)  \,, \qquad
  \bar \epsilon'_{22} = \left( 0 , 0 \right) \,.
 } 
Because $\bar \epsilon_{ip}' \delta \Xi^{ip}$ is by definition $\cQ_H$-exact, we can then write
 \es{QExact}{
		S_m^\text{3d}
		&= -\int\dd^3 x \, \sqrt{g}\,  D_\mu \left(  \bar \epsilon'_{ip}  \gamma^\mu \epsilon^{jp}   J^i{}_j \right) +   \text{(${\cal Q}_H$-exact term)} \,.
  }
One can explicitly check that $ \bar \epsilon'_{ip}  \gamma^\mu \epsilon^{jp}  $ is not a smooth vector field over the entire $S^3_b$---if it were, then, as mentioned above, $S_m^\text{3d}$ would be $\cQ_H$-exact.  The vector field $ \bar \epsilon'_{ip}  \gamma^\mu \epsilon^{jp}  $ fails to be smooth precisely along the $\theta = 0$ circle, but it nevertheless stays bounded everywhere.  To perform the integration in \eqref{QExact}, it is thus useful to split the integration range into two regions:
\begin{equation}\label{RealMassSplit}
	\begin{aligned}
		S_m^\text{3d}
		= -\int_{\theta > \theta_0} \dd^3 x \, \sqrt{g}\,  D_\mu \left(  \bar \epsilon'_{ip}  \gamma^\mu \epsilon^{jp}   J^i{}_j \right)
		 -\int_{\theta < \theta_0} \dd^3 x \, \sqrt{g}\,  D_\mu \left(  \bar \epsilon'_{ip}  \gamma^\mu \epsilon^{jp}   J^i{}_j \right) 
		  +   \text{(${\cal Q}_H$-exact term)} \,,
	\end{aligned}
\end{equation}
for some $\theta_0>0$.  In the limit $\theta_0 \to 0$, the second term vanishes because $ \bar \epsilon'_{ip}  \gamma^\mu \epsilon^{jp}  $ is bounded. Applying Stokes' theorem to the first integral, we obtain
\begin{equation}\label{RealMass}
	S_m^\text{3d}
	= - \lim_{\theta_0 \to 0} \int_\text{$\theta = \theta_0$} d^2 x\,  \sqrt{h}  \, n_\mu   \bar \epsilon'_{ip}  \gamma^\mu \epsilon^{jp}   J^i{}_j + \text{(${\cal Q}_H$-exact term)} \,,
\end{equation}
where $h$ is the determinant of the induced metric on the $\theta = \theta_0$ surface, and $n_\mu$ is the outward pointing unit normal.  Taking $\theta_0 \to 0$ and performing the integral over the angle $\beta$ parameterizing the shrinking circle, the deformation term becomes
 \es{RealMass1dproof}{
	S_m^\text{3d} &= \pi  m r^2 \int_{-\pi}^\pi d \alpha \left[ J^1{}_1 - e^{i \alpha} J^1{}_2 + e^{-i \alpha} J^2{}_1 - J^2{}_2 \right] + \text{(${\cal Q}_H$-exact term)} \\
	&= S_m^\text{1d} + \text{(${\cal Q}_H$-exact term)}  \,.
 }
Hence, $S_m^\text{3d}$ and $S_m^\text{1d}$ differ only by a ${\cal Q}_H$-exact term, which is what we set out to prove.

%%%%%%%%%%%%%%%%%%%%%%%%%%%%%%%%%%%%%
\section{Discussion}
\label{sec:discussion}
%%%%%%%%%%%%%%%%%%%%%%%%%%%%%%%%%%%%%

In this paper we studied the correlation functions of a particular sector of operators in $\cN=4$ QFTs on the squashed sphere.
The main results are summarized as follows:
\begin{itemize}
	\item We formulated $\cN=4$ squashed sphere backgrounds, preserving either a $\su(2|1)\times \psu(1|1)$ or a $\psu(2|2)\times \uu(1)$ superalgebra, and described how to put gauge theories containing vector and hypermultiplets on these backgrounds.
	\item For the $\su(2|1)\times \psu(1|1)$-invariant $\cN=4$ theories, we found two protected one-dimensional sectors, $\cT_H$ and $\cT_C$.  When the ${\cal N} = 4$ theories have Lagrangian descriptions in terms of vector multiplets and hypermultiplets, the $\cT_H$ and $\cT_C$ sectors are related to the Higgs and Coulomb branches of these theories, analogously to the one-dimensional sectors found on the round three-sphere \cite{Dedushenko:2016jxl,Dedushenko:2017avn,Dedushenko:2018icp}. 
	\item For ${\cal N} = 4$ gauge theories with matter hypermultiplets, we used supersymmetric localization to derive an explicit description of the $\cT_H$ sector.   Even though all the intermediate steps were greatly complicated by the introduction of the squashing, the end result is remarkably similar to the round sphere case, and, after appropriate rescalings, the dependence on the squashing parameter $b$ can be removed entirely.
	\item We showed that for any theory on the squashed sphere, irrespective of Lagrangian description, real mass deformations of the full theory are translated into deformations of the one-dimensional theory.  
\end{itemize}

There are various questions we leave open for future research. First of all, it would be interesting to investigate whether there exists a more fundamental reason for the trivial $b$-dependence of correlation functions in the $\cT_H$ sector that we observed in this work.  In particular, it would be interesting to see whether such a trivial $b$-dependence extends beyond the class of theories we studied here and/or to the $\cT_C$ sector.  A possibility is that such a trivial $b$-dependence is required by supersymmetric Ward identities, in which case it is plausible that a general proof would be available.

In this work, we focused on gauge theories with vector and hypermultiplets, but the general framework developed in Section~\ref{sec:1dSectorCohomology} can be applied to a wider range of theories, such as the Chern-Simons matter theories discussed in \cite{Gaiotto:2008ak}, ABJ(M) theory \cite{Aharony:2008ug,Aharony:2008gk}, and generalizations thereof \cite{Imamura:2008dt,Hosomichi:2008jd}. The main difficulty to studying such theories using supersymmetric localization is that at present there is no formulation of such theories where the algebra of the needed supercharges closes off-shell.  Nevertheless, there is no obstruction that we are aware of to obtaining such an off-shell formulation.

As already mentioned in the main text, one can consider other $\cN=4$ supersymmetric backgrounds and investigate whether they admit protected sectors.   While in this paper we studied squashed sphere backgrounds with $\SU(2) \times \UU(1)$ isometry, it would be interesting to see whether such a large isometry group was necessary for preserving the 1d sectors.  One could also investigate, for instance, a squashed sphere with just $\UU(1) \times \UU(1)$ isometry.   Another interesting case is the $S^2 \times S^1$ case discussed in \cite{Panerai:2020boq} for free hypermultiplets coupled to background vector multiplets.

%%%%%%%%%%%%%%%%%%%%%%%%%%%%%%%%%%%%%
\section*{Acknowledgments}
We are grateful to M.~Dedushenko for useful discussions. The work of PB is supported by the STARS-StG grant THEsPIAN, a Francqui Fellowship of the Belgian American Educational Foundation and a Fulbright Fellowship. SSP is supported by the Simons Foundation Grant No. 488653, and by the US NSF under Grant No. 2111977.

%%%%%%%%%%%%%%%%%%%%%%%%%%%%%%%%%%%%%%%%%%%%%%

\appendix

%%%%%%%%%%%%%%%%%%%%%%%%%%%%%%%%%%%%%%%%%%%%%%
\section{Conventions}
\label{app:conventions}
%%%%%%%%%%%%%%%%%%%%%%%%%%%%%%%%%%%%%%%%%%%%%%

In this appendix we collect our conventions and notation. We employ a plethora of indices all of which are summarized in Table \ref{tab:indices} together with their respective meanings.
\begin{table}[!htb]
	\centering
	\begin{tabular}{l|c|c}
		Index type & Range & Meaning \\
		\hline
		\Tstrut$\;\mu,\nu,\rho,\dots$ & $1,2,3$ & Spacetime indices\\
		$\;a,b,c,\dots$ & $1,2,3$ & Tangent bundle indices\\
		$\;i,j,k,\dots$ & $1,2$ & $\SU(2)_H$ indices\\
		$\;p,q,r,\dots$ & $1,2$ & $\SU(2)_C$ indices\\
		$\;\alpha,\beta,\dots$ & $1,2$ & Spinor indices\\
		$\;I,J,K,\dots$ & $1,2,\dots,2n_h$ & Fundamental $\USp(2n_h)$ indices\\
		$\;\wti I,\wti J, \wti K,\dots$ & $1,2,\dots,2\wti n_h$ & Fundamental $\USp(2 \wti n_h)$ indices
	\end{tabular}
	\caption{The various types of indices used in this paper. The labels $n_h$ and $\wti n_h$ denotes the number of hypermultiplets and twisted hypermultiplets respectively.}
	\label{tab:indices}
\end{table}

We will work exclusively in three-dimensional Euclidean signature where we choose the $\gamma$ matrices as follows
\begin{equation}
	\gamma_1 = \sigma_1\,,\qquad \gamma_2 = \sigma_2\,, \qquad \gamma_3 = \sigma_3 \,.
\end{equation}
The Pauli matrices $\sigma_a$ are given by the standard expressions
\begin{equation}
	\sigma_1 = \begin{pmatrix}
		0 & 1 \\ 1 & 0
	\end{pmatrix}\,,\qquad \sigma_2 = \begin{pmatrix}
		0 & -i\\ i & 0
	\end{pmatrix}\,,\qquad \sigma_3 = \begin{pmatrix}
		1 & 0\\ 0 & -1
	\end{pmatrix}\,.
\end{equation}
The charge conjugation matrix $C= i \sigma_2$ and the complex conjugation matrix $B=\alpha C^T$ are defined such that
\begin{equation}
	\gamma_a^T = -C\gamma_a C^{-1}\,,\qquad \gamma_a^* = B\gamma_a B^{-1}\,.
\end{equation}
In Euclidean three-dimensional space one cannot consistently define a reality condition to define Majorana spinors.  We will introduce the following `conjugate' spinor 
\begin{equation}
	\tilde{\lambda}= \lambda^T C\,,
\end{equation}
which after Wick rotation to Lorentzian signature becomes the Majorana conjugate of the Lorentzian spinor. Both the spinorial and $\SU(2)_{H/C}$ indices are raised and lowered with the $\SU(2)$-invariant tensor $\varepsilon^{ij}$, $\varepsilon^{pq}$ or $\varepsilon^{\alpha\beta}$. We use the NW-SE convention for contracting both spinorial and $\SU(2)_{H/C}$ indices, and hence we have
\begin{equation}
	\varepsilon_{ij}\varepsilon^{jk} = -\delta_i^k\,,\qquad \varepsilon_{12}=\varepsilon^{12}=1\,.
\end{equation}
The symplectic indices are raised and lowered with the symplectic form $\varepsilon_{IJ}$ or $\varepsilon_{\tilde I \tilde J}$ following the same NW-SE convention.

%%%%%%%%%%%%%%%%%%%%%%%%%%%%%%%%%%%%%%%%%%%
\section{3d $\cN=4$ conformal supergravity}
\label{app:confSugra}
%%%%%%%%%%%%%%%%%%%%%%%%%%%%%%%%%%%%%%%%%%%

In this Appendix~we collect some useful facts about $\cN=4$ conformal supergravity in 3d. We use the conventions of \cite{Banerjee:2015uee}, and we will summarize their results for the supersymmetry transformations and conformal actions for hypermultiplets and vector multiplets. In addition, we give some more details on how one can gauge fix the conformal action for the vector multiplets using a compensator vector multiplet in order to obtain the Yang-Mills action on the squashed sphere.

\subsection{Supersymmetry variations}

The field content of the 3d $\cN=4$ Weyl multiplet is given by
\begin{equation}
	\begin{aligned}
		{\rm Bosonic:}&\quad e_\mu{}^a\,,\quad  b_\mu\,,\quad V_\mu{}^i{}_j \,,\quad \wti V_\mu{}^p{}_q\,,\quad C\,,\quad D\,,\\
		{\rm Fermionic:}&\quad \psi_\mu^{ip}\,,\quad \chi^{ip}\,.
	\end{aligned}
\end{equation}
The Poincar\'e and conformal supersymmetry transformations of the fields are\footnote{Note that the sign of the field strengths of the R-symmetry background fields in the supersymmetry variation for $\chi$ are opposite to those in \cite{Banerjee:2015uee}. One can check that this is indeed the correct sign in our conventions by checking the supersymmetry algebra on the R-symmetry gauge fields, i.e.
	\begin{equation}\label{delta2VAgain2}
		\comm{\delta_1}{\delta_2} V_{\mu}{}^i{}_{j} =   \xi^a G_{a\mu}{}^i{}_ j  -   2\partial_\mu v^i{}_j 
		-V_\mu{}^i{}_k v^k{}_j + V_\mu{}^k{}_j v^i{}_k  - \text{trace}\,.
	\end{equation}
	The first term gives the correct covariant general coordinate transformation, and it would have been absent with the original sign.
}
\begin{align}
	\delta\psi_\mu^{ip} &= 2\cD_\mu \epsilon^{ip} - \gamma_\mu \eta^{ip}\,,\\
	\delta\chi^{ip} &= 2\slashed{D}C \epsilon^{ip} + D\epsilon^{ip} - \f12 \slashed{\wti G}^p{}_q \epsilon^{iq} + \f12 \slashed{G}^i{}_j \epsilon^{jp} + 2C \eta^{ip}\,,\\
	\delta e_\mu{}^a &= \bar{\epsilon}_{ip}\gamma^a\psi_\mu^{ip}\,,\\
	\delta b_\mu &= \f12 \bar{\epsilon}_{ip}\phi_\mu^{ip} - \f12 \bar{\eta}_{ip}\psi_\mu^{ip} + \lambda_K^a e_{\mu a}\,,\\
	\delta \cV_\mu{}^i{}_j &= \bar{\epsilon}_{jp}\phi_\mu{}^{ip}-2C \bar{\epsilon}_{jp}\psi_\mu{}^{ip} - \bar{\epsilon}_{jp}\gamma_\mu\chi^{ip} + \bar{\eta}_{jp}\psi_\mu{}^{ip} - {\rm trace}\,,\\
	\delta \cA_\mu{}^p{}_q &= \bar{\epsilon}_{iq}\phi_\mu{}^{ip}+2C \bar{\epsilon}_{iq}\psi_\mu{}^{ip} + \bar{\epsilon}_{iq}\gamma_\mu\chi^{ip} + \bar{\eta}_{iq}\psi_\mu{}^{ip} - {\rm trace}\,,\\
	\delta C &= \f12 \bar{\epsilon}_{ip}\chi^{ip}\,,\\
	\delta D &= \bar{\epsilon}_{ip} \slashed{D}\chi^{ip} - \bar{\eta}_{ip}\chi^{ip}\,.
\end{align}
In these equations, $G$ and $\wti G$ denote the field strengths of $V$ and $\wti V$, respectively, and the covariant derivative of $\epsilon$ is defined as
\begin{equation}
	\cD_\mu \epsilon^{ip} = \left( \partial_\mu + \f14 \omega_\mu^{ab}\gamma_{ab}+\f12 b_\mu \right)\epsilon^{ip} + \f12 V_\mu{}^i{}_j \epsilon^{jp} + \f12 \wti V_\mu{}^p{}_q \epsilon^{iq}\,.
\end{equation}
Further, from the curvature constraints on finds
\begin{equation}
	f_\mu{}^a = R_\mu{}^a - \f14 e_\mu{}^a R\,.
\end{equation}

In 3d $\cN=4$ supergravity, vector and twisted vector multiplets have the following components
\begin{center}
	\begin{tabular}{rcc}
		& $\qquad$Vector multiplet $\cV$$\qquad$ & Twisted vector multiplet $\wti\cV$\\
		\Tstrut Bosonic:&  $L^p{}_q\,, \quad Y^i{}_j\,,\quad A_\mu\,,$ & $\wti{L}^i{}_j\,, \quad \wti{Y}^p{}_q\,,\quad \wti{A}_\mu\,,$ \\
		Fermionic:& $\Omega^{ip}\,$ &  $\wti\Omega^{ip}\,.$
	\end{tabular}
\end{center}
Their supersymmetry variations are given by
\begin{align}\label{SUSYTvec}
	\begin{split}
		\delta\Omega^{ip} &= \slashed{\cD}L^p{}_q\epsilon^{iq}-\f12 F_{ab}\gamma^{ab}\epsilon^{ip}+ Y^i{}_j\epsilon^{jp}+C L^p{}_q\epsilon^{iq} + L^p{}_q\eta^{iq} -\f12 \comm{L^p{}_q}{L^q{}_r}\epsilon^{ir} \,,\\
		\delta L^p{}_q &= 2\bar{\epsilon}_{iq}\Omega^{ip}-\delta^p{}_q\bar{\epsilon}_{ir}\Omega^{ir}\,,\\
		\delta A_\mu &= \bar{\epsilon}_{ip}\gamma_\mu\Omega^{ip} + L^p{}_q \bar{\epsilon}_{ip}\psi_\mu{}^{iq}\,,\\
		\delta Y^i{}_j &= 2\bar{\epsilon}_{jp}\slashed{D}\Omega^{ip}-L^p{}_q\bar{\epsilon}_{jp}\chi^{iq} - 2 C \bar{\epsilon}_{jp}\Omega^{ip} - \bar{\eta}_{jp}\Omega^{ip} +  \bar{\epsilon}_{jp}\comm{\Omega^{iq}}{L^p{}_q} - {\rm trace}\,,
	\end{split}
\end{align}
and
\begin{align}\label{SUSYTtvec}
	\begin{split}
		\delta\wti\Omega^{ip} &= \slashed{\cD}\wti{L}^i{}_j\epsilon^{jp}-\f12 \wti F_{ab}\gamma^{ab}\epsilon^{ip}+ \wti{Y}^p{}_q\epsilon^{iq} - C \wti{L}^i{}_j\epsilon^{jp} + \wti{L}^i{}_j\eta^{jp}-\f12 \comm{\wti L^i{}_j}{\wti L^j{}_k}\epsilon^{kp}\,,\\
		\delta \wti{L}^i{}_j &= 2\bar{\epsilon}_{jp}\wti\Omega^{ip}-\delta^i{}_j\bar{\epsilon}_{kp}\wti\Omega^{kp}\,,\\
		\delta \wti{A}_\mu &= \bar{\epsilon}_{ip}\gamma_\mu\wti\Omega^{ip} + \wti{L}^i{}_j \bar{\epsilon}_{ip}\psi_\mu{}^{jp}\,,\\
		\delta \wti{Y}^p{}_q &= 2\bar{\epsilon}_{iq}\slashed{D}\wti\Omega^{ip}+\wti{L}^i{}_j\bar{\epsilon}_{jq}\chi^{ip} + 2 C \bar{\epsilon}_{iq}\wti\Omega^{ip} - \bar{\eta}_{iq}\wti\Omega^{ip} +  \bar{\epsilon}_{iq}\comm{\wti\Omega^{jp}}{\wti L^i{}_j} - {\rm trace}\,,
	\end{split}
\end{align}
where $F$ and $\wti F$ are the field strengths of $A$ and $\wti A$ respectively, and the covariant derivatives on $L$ and $\wti{L}$ are defined as
\es{CovDerL}{
	\cD_\mu L^p{}_q &= \partial_\mu L^p{}_q +\f12 \wti V_{\mu}{}^p{}_r L^r{}_q - \f12 \wti V_{\mu}{}^r{}_q L^p{}_r\,,\\
	\cD_\mu \wti{L}^i{}_j &= \partial_\mu \wti{L}^i{}_j +\f12 V_{\mu}{}^i{}_k \wti{L}^k{}_j - \f12 V_{\mu}{}^k{}_j \wti{L}^i{}_k\,.
}

Finally we also consider hypermultiplets and their twisted relatives. Their field content is given by
\begin{center}
	\begin{tabular}{rcc}
		& $\qquad$ hypermultiplet $\cH^I$ $\qquad$ & Twisted hypermultiplet $\wti\cH^{\wti I}$\\
		\Tstrut Bosonic:&  $z_i{}^I\,,$ & $\wti z_p{}^{\wti I}\,,$ \\
		Fermionic:& $\zeta^{Ip}\,$ &  $\wti\zeta^{\wti I i}\,,$
	\end{tabular}
\end{center}
A set of $n_h$ hypermultiplets consists of $2n_h$ scalars $z_i{}^I$ and $2n_h$ fermions $\zeta^{Ip}$, where $I=1,\dots,2n_h$. Similarly, a $\wti n_h$ twisted hypermultiplets consists of scalars $\wti{z}_p{}^{\wti I}$ and fermions $\wti{\zeta}^{\wti I i}$, with $\wti I = 1,\dots , 2\wti n_h$ but they transform in the opposite way under the $\SU(2)_H\times \SU(2)_C$ R-symmetry. The supersymmetry variations are given by
\begin{align}\label{hypvars}
	\delta z_i{}^I &= 2\bar{\epsilon}_{ip}\zeta^{Ip}\,,\\
	\delta \zeta^{Ip} &= \slashed{\cD}z_i{}^I \epsilon^{ip} - \f12 C z_i{}^I \epsilon^{ip} + \f12 z_i{}^I\eta^{ip}\,,\\
	\delta \wti{z}_p{}^{\wti I} &= 2\bar{\epsilon}_{ip}\wti{\zeta}^{\wti I i}\,,\\
	\delta \wti{\zeta}^{\wti I i} &= \slashed{\cD}\wti{z}_p{}^{\wti I} \epsilon^{ip} + \f12 C \wti{z}_p{}^{\wti I} \epsilon^{ip} + \f12 \wti{z}_p{}^{\wti I}\eta^{ip}\,,
\end{align}
where the covariant derivatives are given by
\begin{align}\label{covDz}
	\cD_a z_i{}^I &= \partial_a z_i{}^I - \f12  z_j^I V_a{}^j{}_i\,,\\
	\cD_a \wti{z}_p{}^{\wti I} &= \partial_a \wti{z}_p{}^{\wti I} - \f12  \wti{z}_q{}^{\wti I}\wti V_a{}^q{}_p\,.
\end{align}
The fields in the hypermultiplet transform in the fundamental representation of $\USp(2n_h)$. When we consider hypermultiplets coupled to a background vector multiplet gauging a subgroup of $\USp(2 n_h)$, the fields in the hypermultiplet transform in the appropriate representation $\cR$ of the gauge group $G$. In this case an additional term proportional to the gauge field has to be added to the covariant derivative \eqref{covDz}, namely
\begin{align}\label{covDzGauge}
	\cD_a z_i{}^I &= \partial_a z_i{}^I - \f12  z_j^I V_a{}^j{}_i - i A_a{}^I{}_J z_i{}^J\,,\\
	\cD_a \wti{z}_p{}^{\wti I} &= \partial_a \wti{z}_p{}^{\wti I} - \f12  \wti{z}_q{}^{\wti I}\wti V_a{}^q{}_p- i \wti A_a{}^{\wti I}{}_{\wti J} \wti z_p{}^{\wti J}\,,
\end{align}
where we also wrote the analogous equation for the twisted hypermultiplets gauged under a twisted vector multiplet.  Finally, in the main text we introduced the conserved multiplet $\cJ = \star \wti\cV$ and its twisted version $\wti\cJ = \star \cV$ in Eqs.~\eqref{Ident} and \eqref{Identtwisted}. The supersymmetry variations for these multiplets can be obtained from those of the twisted vector and vector multiplet, respectively, and are given by
\begin{equation}\label{currentSUSY}
	\begin{aligned}
		\delta J^i{}_j &= 2\bar{\epsilon}_{jp}\Xi^{ip}-\delta_j^i \bar{\epsilon}_{kp}\Xi^{kp}\,,\\
		\delta\Xi^{ip} &= \slashed{\cD}J^i{}_j\epsilon^{jp}- i j^c\gamma_c \epsilon^{ip} + K^p{}_q \epsilon^{iq}-C J^i{}_j \epsilon^{jp} + J^i{}_j \eta^{jp}\,,\\
		\delta j^a &= - i\bar{\epsilon}_{ip}\gamma^{ac}\cD_c \Xi^{ip}-i\bar{\eta}_{ip}\gamma^a\Xi^{ip}\,,\\
		\delta K^p{}_q &= 2\bar{\epsilon}_{iq}\slashed{\cD}\Xi^{ip} + J^i{}_j\bar{\epsilon}_{iq}\chi^{jp} + 2C\bar{\epsilon}_{iq}\Xi^{ip} - \bar{\eta}_{iq}\Xi^{ip} - {\rm trace}\,,
	\end{aligned}
\end{equation}
and
\begin{equation}\label{currentTSUSY}
	\begin{aligned}
		\delta \wti J^p{}_q &= 2\bar{\epsilon}_{iq}\wti\Xi^{ip}-\delta_q^p \bar{\epsilon}_{iq}\Xi^{iq}\,,\\
		\delta\wti\Xi^{ip} &= \slashed{\cD}\wti J^p{}_q\epsilon^{iq}- i \wti j^c\gamma_c \epsilon^{ip} +\wti K^i{}_j \epsilon^{jp}-C \wti J^p{}_q \epsilon^{iq} + \wti J^p{}_q \eta^{iq}\,,\\
		\delta \wti j^a &= - i\bar{\epsilon}_{ip}\gamma^{ac}\cD_c \wti \Xi^{ip}-i\bar{\eta}_{ip}\gamma^a\wti\Xi^{ip}\,,\\
		\delta\wti K^i{}_j &= 2\bar{\epsilon}_{jp}\slashed{\cD}\wti \Xi^{ip} + \wti J^p{}_q\bar{\epsilon}_{jp}\chi^{iq} + 2C\bar{\epsilon}_{jp}\wti \Xi^{ip} - \bar{\eta}_{jp}\wti \Xi^{ip} - {\rm trace}\,.
	\end{aligned}
\end{equation}

\subsection{Lagrangians and gauge fixing}
\label{LAGRANGIANS}

The Lagrangian of a collection of $n_h$ hypermultiplets charged under a vector multiplet is given by 
\begin{equation}
	\begin{aligned}
		S_{\rm hyp}[\cH^I,\cV] =& -\f12\int \sqrt{g} \varepsilon_{IJ}\Bigg( \cD_\mu z_i{}^I\cD^\mu z^{iJ} + \f14 z_i{}^I z^{iJ}\left( \f12 R - D + C^2 \right) \\
		&\qquad\qquad\qquad+ \f{1}{2} z_i{}^I L^p{}_q{}^J{}_K L^q{}_p{}^K{}_L z^{iL} + i\,z_i{}^I Y^{i}{}_j{}^J{}_K z^{jK}\\
		&\qquad\qquad\qquad + i \overline\zeta^{pI}\slashed{\cD}\zeta_p^J + i \overline\zeta_p^I L^{p}{}_q{}^J{}_K \zeta^{qK} + z_i{}^I \overline\Omega^{ip}\zeta{_p^J}\Bigg)\,.
	\end{aligned}
\end{equation}
Similarly, the Lagrangian for $\wti{n}_h$ charged twisted hypermultiplets is given by
\begin{equation}
	\begin{aligned}
		S_{\rm t. hyp}[\wti\cH^{\wti I},\wti\cV] =& -\f12\int \sqrt{g} \wti \varepsilon_{\wti I \wti J} \Bigg( \cD_\mu \wti{z}_p{}^{\wti I}\cD^\mu \wti{z}^{p\wti J} + \f14 \wti{z}_p{}^{\wti I} \wti{z}^{p\wti J}\left( \f12 R + D + C^2 \right) \\
		&\qquad\qquad\qquad+\f12 \wti{z}_p{}^{\wti I} \wti L^i{}_j{}^{\wti J}{}_{\wti K} \wti L^j{}_i{}^{\wti K}{}_{\wti L} \wti{z}^{p\wti L} + i\,\wti{z}_p{}^{\wti I} \wti Y^{p}{}_q{}^{\wti J}{}_{\wti K} \wti{z}^{p\wti K}\\
		&\qquad\qquad\qquad + i \overline{\wti\zeta}{}^{i\wti I}\slashed{\cD}\wti\zeta_i^{\wti J} + i \overline{\wti\zeta}_i{}^{\wti I} \wti L^{i}{}_j{}^{\wti J}{}_{\wti K} \wti\zeta^{j\wti K} + \wti z_i{}^{\wti I} \overline{\wti\Omega}{}^{ip}\wti\zeta_p^{\wti J}\Bigg)\,.
	\end{aligned}
\end{equation}
Here, we introduced the skew-symmetric $\USp(n_h)$-invariant tensor $\varepsilon_{IJ}$ and $\USp(\wti n_h)$-invariant tensor $\wti \varepsilon_{\wti I\wti J}$. For concreteness we can take these rank-two tensors to be
\begin{equation}
	\varepsilon_{IJ} = \wti\varepsilon_{\wti I \wti J} = \begin{pmatrix}
		i \sigma_2 & 0 & 0 & \cdots \\
		0 & i \sigma_2 & 0 & \cdots \\
		0 & 0 & i \sigma_2 & \cdots \\
		\vdots & \vdots & \vdots & \ddots \end{pmatrix} \,.
\end{equation}
These action are invariant under the full $\osp(4|4)$ algebra. Hence the action for hypermultiplets on the squashed sphere is simply obtained from the flat space action by covariantizing all derivatives and inserting the additional mass terms given by the second term on the first line. 

For the vector multiplets we will be interested in a Yang-Mills kinetic term. The Yang-Mills action, however, does not preserve the full superconformal group, and it cannot simply be inferred from the flat space action. To construct the non-conformal YM action for the vector multiplets we will use the tools of superconformal tensor calculus (see \cite{Freedman:2012zz,Lauria:2020rhc} for a review). The strategy we will follow is to start with conformal supergravity coupled to a set of vector multiplets and gauge fix the unwanted conformal symmetries by adding an additional abelian compensator vector multiplet. Starting from a conformal action for $n_v+1$ vector multiplets we can then obtain the non-conformal action for $n_v$ vector multiplets by fixing the background values of the additional compensator vector multiplet.

Let us very briefly describe the various steps in this process. Starting from conformal supergravity we gauge fix the special conformal transformations by setting $b_\mu=0$. Since, all elementary fields, except $b_\mu$ itself, transform trivially under special conformal transformations, this gauge is automatically preserved. In order to fix the remaining symmetries we introduce an additional compensator vector multiplet $\cV_0$. To fix the $\SU(2)_C$ R-symmetry to its Cartan subalgebra we set 
\begin{equation}
	L_0{}^p{}_q = m (\sigma_3)^p{}_q\,.
\end{equation}
In addition, we set $\Omega_0^{ip}=0$, which fixes the special conformal supersymmetry transformations. However, to preserve this gauge choice we have to accompany every Poincar\'e supersymmetry transformation by a conformal supersymmetry transformation with the following field dependent parameter
\begin{equation}\label{etafix}
	\eta^{ip} = \f{1}{2m}\slashed{F}_0 (\sigma_3)^p{}_q\epsilon^{iq} -  \f{1}{m}(\sigma_3)^p{}_qY_0{}^i{}_j \epsilon^{jq} -  C\epsilon^{ip}\,. 
\end{equation}
The supergravity theory thus obtained will be a $\SU(2)_H\times \UU(1)_C$ gauged $\cN=4$ Poincar\'e supergravity theory, and the Poincar\'e supersymmetry transformations act as $\delta = \delta_\epsilon + \delta_\eta$ where $\eta$ is determined by \eqref{etafix}. In a similar fashion we can add an additional compensator twisted vector multiplet to gauge fix the $\SU(2)_H$ symmetry to its Cartan. This will be necessary when we want to construct a YM action for twisted vector multiplets but we will not discuss this in detail.

Having discussed the gauge fixing, we can now continue to construct the YM action. To do so we start from the following conformal action for $n_v+1$ vector multiplets \cite{Banerjee:2015uee},
\begin{equation}\label{vecConf}
	\begin{aligned}
		S_{\rm vec} = \f{1}{g_{YM}^2} \int \sqrt{g} \cF_{\Sigma\Lambda}  \Bigg(& \cD_\mu L^{p}{}_q{}^\Sigma \cD^\mu L^{q}{}_p{}^\Lambda + L^{p}{}_q{}^\Sigma L^{q}{}_p{}^\Lambda \left( \f12 R + D + C^2 \right) \\
		&- F_{\mu\nu}{}^\Sigma F^{\mu\nu\Lambda} + Y^{i}{}_j{}^\Sigma Y^{j}{}_i{}^\Lambda \Bigg)\,,
	\end{aligned}
\end{equation}
where $\Sigma,\Lambda=0,1,\cdots,n_v$. The zeroth vector multiplet corresponds to the compensating vector multiplet and the function $\cF_{\Sigma\Lambda}(L) = \f{\partial^2\cF}{\partial L^\Sigma \partial L^\Lambda}$ is a function of the vector multiplet scalars $L^p{}_q$ that can be obtained as the second derivative of the prepotential $\cF$, which in turn encodes the geometry of the scalar manifold. The dependence of $\cF$ on the abelian compensating multiplet is fully determined and we can write it as
\begin{equation}
	\cF=2m^2\frac{\cF_{AB}L{}^p{}_q{}^A L^q{}_p{}^B-L_0{}^r{}_sL_0{}^s{}_r}{L_0{}^r{}_sL_0{}^s{}_r}\,,
\end{equation}
where $A,B=1,\dots , n_v$ and the minus sign reflects the fact that $L_0$ belongs to a compensator vector multiplet. Using this expression for the prepotential we can determine the functions $\cF_{\Sigma,\Lambda}$ as follows,
\begin{align}
	\cF_{AB} &= \cF_{AB}\,,\\
	\cF_{A0} &= -\f{1}{2m^2}\cF_{AB} L^p{}_q{}^B L_0{}^q{}_p\,,\\
	\cF_{00} &= -\f{1}{4m^2}\cF_{AB}\left( L^p{}_q{}^A L^q{}_p{}^B - \f{3}{2m^2} L^p{}_q{}^AL_0{}^q{}_pL^r{}_s{}^BL_0{}^s{}_r\right)\,.
\end{align}
Substituting these expressions in the conformal action \eqref{vecConf} we obtain the non-conformal Yang-Mills action
\begin{align}\label{LYM}
	S_{\rm YM}[\cV] = & \f{1}{g_{YM}^2} \int \sqrt{g} \cF_{AB}  \Bigg( \cD_\mu L^{p}{}_q{}^A \cD^\mu L^{q}{}_p{}^B + F_{\mu\nu}{}^A F^{\mu\nu B} - Y^{i}{}_j{}^A Y^{j}{}_i{}^B \nn\\
	&+ L^{p}{}_q{}^A L^{q}{}_p{}^B \left[\f12 \left( \f12 R + D + C^2 \right) + \f{1}{4 m^2} \left( F_{0\,\mu\nu}{} F_0^{\mu\nu} + Y_0{}^{i}{}_j{} Y_0{}^{j}{}_i{} \right) \right]\\
	&- \f{1}{m^2} \left(L^{p}{}_q{}^A L_0{}^q{}_p\right)\left( L^{r}{}_s{}^BL_0{}^s{}_r\right) \left[\f14 \left( \f12 R + D + C^2 \right) - \f{3}{8m^2} \left( F_{0\,\mu\nu}{} F_0^{\mu\nu} + Y_0{}^{i}{}_j{} Y_0{}^{j}{}_i{} \right) \right]\nn\\
	&+\f{1}{2m^2} \left(Y^{i}{}_j{}^A Y_0{}^{j}{}_i\right) \left(L^{p}{}_q{}^B L_0{}^q{}_p\right) \Bigg)\,,\nn
\end{align}
Finally, we can now insert the values for the compensator background vector multiplet for the respective background which reproduces the bosonic part of the Yang-Mills actions \eqref{YMTerm}--\eqref{YMTerm2} quoted in the main text.

In this work we are mainly interested in theories built out of vector and hypermultiplets. However, one can also consider theories with twisted hypermultiplets and twisted vector multiplets. To write down a twisted Yang-Mills term for dynamical twisted vector multiplets one can proceed analogous as for the untwisted vector multiplet. In addition to the gauge fixing procedure introduced above one needs to add an additional compensator twisted vector multiplet to gauge fix the $\SU(2)_C$ to its Cartan. After this one can start from an analogous conformal action as \eqref{vecConf} and substitute the background values of the compensator twisted vector multiplet to obtain the twisted Yang-Mills action. We will not go through this procedure in detail but simply state the resulting action:
\begin{equation}\label{LYMt}
	\begin{aligned}
		S_{\rm YM}[\wti \cV] = & \f{1}{\tilde g_{YM}^2} \int \sqrt{g} \wti\cF_{AB}  \Bigg( \cD_\mu \wti L^{i}{}_j{}^A \cD^\mu \wti L^{j}{}_i{}^B + \wti F_{\mu\nu}{}^A \wti F^{\mu\nu B} -\wti Y^{p}{}_q{}^A \wti Y^{q}{}_p{}^B \\
		&+ \wti L^{i}{}_j{}^A \wti L^{j}{}_i{}^B \left[\f12 \left( \f12 R - D + C^2 \right) + \f{1}{4 \wti m^2} \left( \wti F_{0\,\mu\nu}{} \wti F_0^{\mu\nu} +\wti Y_0{}^{p}{}_q{} \wti Y_0{}^{q}{}_p{} \right) \right]\\
		&- \f{1}{\wti m^2} \left(\wti L^{i}{}_j{}^A \wti L_0{}^j{}_i\right)\left( \wti L^{k}{}_l{}^B \wti L_0{}^l{}_k\right) \left[\f14 \left( \f12 R - D + C^2 \right) - \f{3}{8\wti m^2} \left( \wti F_{0\,\mu\nu}{}\wti F_0^{\mu\nu} + \wti Y_0{}^{p}{}_q \wti Y_0{}^{q}{}_p \right) \right]\\
		&+\f{1}{2\wti m^2} \left(\wti Y^{p}{}_q{}^A \wti Y_0{}^{q}{}_p\right) \left(\wti L^{i}{}_j{}^B \wti L_0{}^j{}_i\right) \Bigg)\,,
	\end{aligned}
\end{equation}
which can equivalently be obtained from the untwisted action \eqref{LYMt} through the mirror map \eqref{MirrorMap}.

%%%%%%%%%%%%%%%%%%%%%%%%%%%%%
\section{Supersymmetry algebra}
\label{app:susyAlgebra}
%%%%%%%%%%%%%%%%%%%%%%%%%%%%%

In this final appendix we discuss in some detail the $\cN=4$ superconformal algebra and how its generators act on local operators. Although the main application in this paper is to non-conformal theories for which the conformal symmetry is broken, it is nonetheless very useful to study the action of the full superconformal algebra. In particular, the backgrounds introduced in the main text all preserve some $\cN=4$ subalgebra of the $\cN=4$ superconformal algebra and hence the action of this superalgebra can be inferred from the action of the full superconformal algebra. 

The three-dimensional $\cN=4$ superconformal group is given by $\OSp(4|4)$ which contains the maximal bosonic subgroup $\SO(4)_R \times \USp(4)$. The spacetime symmetries are generated by translation, special conformal transformations, Lorentz transformations and dilatations with generators
\begin{equation}
	\bP_a\,,\qquad\qquad \bK_a\,\qquad\qquad \bM_{ab}\,,\qquad\qquad \bD\,.
\end{equation}
The dilatation operator acts on a field with Weyl weight $w$ as
\begin{equation}
	\bD\phi = w\phi \,,
\end{equation}
while the generators of Lorentz transformations acts on fermions as 
\begin{equation}
	\bM_{ab}\chi = \f12 \gamma_{ab}\chi\,.
\end{equation} 
In addition to the conformal symmetries, the bosonic part of the algebra contains the R-symmetry $\SO(4)_R \simeq \SU(2)_H\times \SU(2)_C$. We denote its generators by $\bH^i{}_j$ and $\bC^p{}_q$, respectively, and use conventions where they act on the fundamental representation as
\begin{equation}\label{Raction}
	\bH^i{}_j\phi^k = \delta_j^k\phi^i - \f12 \delta^i_j \phi^k\,,\qquad \bC^p{}_q\phi^r = \delta_q^r\phi^p - \f12 \delta^p_q \phi^r\,.
\end{equation}
With these definitions, the covariant derivative with respect to the standard superconformal gauge fields becomes
\begin{equation}
	\cD_\mu \equiv \partial_\mu + \f12 \omega_\mu^{ab}\bM_{ab} - b_\mu \bD - f_\mu{}^a \bK_a + \f12 V_\mu{}^i{}_j \bH^j{}_i + \f12 \wti V_\mu{}^p{}_q \bC^q{}_p + {\rm fermions}\,. 
\end{equation}
By adding 8 Poincar\'e supercharges $\cQ_\alpha^{ip}$, $\wti{\cQ}_{\alpha\,ip}$ and eight conformal supercharges $\cS^\alpha_{ip}$, $\wti{\cS}^{\alpha\,ip}$ we obtain the full $\cN=4$ superconformal algebra. The supersymmetry generators act on an operator $\cO$ (without spacetime indices) as
\begin{equation}
	\delta\cO = \left(\tilde{\epsilon}^{ip}\cQ_{ip} + \tilde{\eta}^{ip}\cS_{ip}\right)\cO \,.
\end{equation}
The supersymmetry parameters $\epsilon$ and $\eta$ are anti-commuting hence the commutator of two variations is related to the anti-commutator of two supercharges and given by
\begin{equation}\label{genQcomm}
	\comm{\delta_1}{\delta_2} = \delta_{\rm cgct}(\xi) + \f12 \lambda^{ab}\bM_{ab} + \lambda_K^a \bK_a + \lambda_D \bD + v^i{}_j \bH^j{}_i + \wti v^p{}_q \bC^q{}_p \,,
\end{equation}
where $\delta_{\rm cgct}(\xi)$ is a covariant general coordinate transformation which is given by $\delta_{\rm cgct}(\xi)=\xi^\mu D_\mu$ when acting on a scalar field. The coefficients in \eqref{genQcomm} are given by various spinor bilinears defined in terms of the supersymmetry parameters as follows
\begin{equation}\label{susyparams}
	\begin{aligned}
		\xi^\mu &= 2\tilde{\epsilon}_{2ip}\gamma^\mu \epsilon_{1}^{ip}\,,\\
		\lambda^{ab} &=  \tilde{\epsilon}_{2ip}\gamma^{ab} \eta_{1}^{ip} + \tilde{\eta}_{2ip}\gamma^{ab} \epsilon_{1}^{ip}\,,\\
		\lambda^c_K &= \tilde{\eta}_{2ip}\gamma^c \eta_{1}^{ip} + \f{i}{2}\varepsilon^{abc}\tilde{\epsilon}_{2ip}\left( G_{ab}{}^i{}_j\epsilon_1^{jp}+{\wti G}_{ab}{}^p{}_q\epsilon_1^{iq} \right)\,,\\
		\lambda_D &= - \tilde{\epsilon}_{2ip}\eta_{1}^{ip} + \tilde{\eta}^{ip}_2\epsilon_{1ip} \,,\\
		v^i{}_j &= -\tilde{\epsilon}_{2jp} \eta_{1}^{ip} - \tilde{\eta}_{2jp} \epsilon_{1}^{ip} +2 C \tilde{\epsilon}_{2jp} \epsilon_{1}^{ip} - {\rm trace}\,,\\
		\wti v^p{}_q &= -\tilde{\epsilon}_{2iq} \eta_{1}^{ip} - \tilde{\eta}_{2iq} \epsilon_{1}^{ip} - 2 C \tilde{\epsilon}_{2iq} \epsilon_{1}^{ip} - {\rm trace}\,.
	\end{aligned}
\end{equation}
Our solutions preserve either the superalgebra $\su(2|1)\times \psu(1|1)$ or $\psu(2|2)\times \uu(1)$ and therefore the parameters \eqref{susyparams} should be restricted to lie in this subalgebra. In addition, as described in the previous appendix, whenever we perform a supersymmetry transformation in our non-conformal theory it should be accompanied by a conformal supersymmetry transformation with field dependent supersymmetry parameter \eqref{etafix}.

\subsection{Central extensions of the supersymmetry algebra}

When the theory contains flavor symmetries $G_F$ or gauge symmetries $G$, we can couple our background to background vector and twisted vector multiplets valued in the Cartan of $G_F$ or $G$, respectively. Adding such background multiplets corresponds to adding real masses or FI terms, respectively. In an $\cN=4$ superconformal theory, these additional parameters would break part of the superconformal symmetries. However, they preserve the full supersymmetry algebras of the squashed sphere backgrounds introduced in Section~\ref{sec:background}, and, as already mentioned in the main text, they correspond to central extensions of it.

For our purposes it will suffice to consider abelian symmetries only, in which case we will denote the generators of the global and gauge symmetries by $\bQ$ and $\wti{\bQ}$, respectively. These generators are defined such that in the abelian case they act on a field $\phi$ of charge $+1$ as
\begin{equation}
	\bQ \phi = \wti{\bQ}\phi = \phi\,.
\end{equation}
In the presence of background (twisted) vector multiplets, we have to modify the covariant derivative to include a connection term involving the abelian symmetry generator $\bQ$ or its twisted analog $\wti{\bQ}$,
\begin{equation}\label{Dcov}
	\begin{aligned}
		\cD_\mu = \partial_\mu + \frac 12 \omega_\mu{}^{ab}  \bM_{ab} - b_\mu \bD -  f_\mu{}^a \bK_a + \frac 12 V_\mu{}^i{}_j &\bH^j{}_i + \frac 12 \wti V_\mu{}^p{}_q \bC^q{}_p 
		\\
		&- i A_\mu \bQ - i\wti{A}_\mu \wti{\bQ} + \text{fermions} \,.
	\end{aligned}
\end{equation}
In addition to modifying the covariant derivative, this has the effect of centrally extending the supersymmetry algebra. This effect can be observed as additional gauge transformations in the commutator of two supersymmetries:
\begin{align}\label{QcommQ}
	\comm{\delta_1}{\delta_2} &=  \delta_\text{cgct}(\xi) + \frac 12 \lambda^{ab} {\bf M}_{ab} + \lambda_K^a {\bf K}_a + \lambda_D {\bf D}
	+ v^i{}_j {\bf H}^j{}_i + \wti v^p{}_q {\bf C}^q{}_p + \lambda_\bQ {\bf Q} + \lambda_{\tilde \bQ} \tilde {\bf Q}  \,.
\end{align}
The parameters $\lambda_{\bQ}$ and $\lambda_{\wti \bQ}$ can be read off from the commutator acting on vector in an abelian (twisted) vector multiplet,
\begin{align}
	\comm{\delta_1}{\delta_2} A_\mu &= \xi^a F_{a\mu} + \partial_\mu\left( 2L^p{}_q\overline\epsilon_{2\,ip}\epsilon_1{}^{iq} \right)\,,\\
	\comm{\delta_1}{\delta_2} \wti A_\mu &= \xi^a \wti F_{a\mu} + \partial_\mu\left( 2\wti L^i{}_j\overline\epsilon_{2\,ip}\epsilon_1{}^{jp} \right)\,.
\end{align}
The first term on the right-hand sides of these equations represents a covariant general coordinate transformation, while the second terms represent a gauge transformation with parameters
\begin{equation}\label{lQ}
	\lambda_Q = 2 L^p{}_q \tilde\epsilon_{2ip}   \epsilon_1{}^{iq} \,, \qquad
	\lambda_{\wti{Q}} = 2 \wti{L}^i{}_j  \tilde{\epsilon}_{2ip}   \epsilon_1{}^{jp}  \,.
\end{equation}
When a hypermultiplet or twisted hypermultiplet is charged under a vector multiplet with gauge charges $q_I$ for the fields $z_i{}^I$ and $\zeta^{Ip}$ (and equivalently, twisted gauge charges $\wti q_{\wti I}$ for $\wti z_p{}^{\wti I}$ and $\wti\zeta^{i\wti I}$) the supersymmetry transformations \eqref{hypvars} are modified to include a gauge transformation: 
\begin{equation}\label{hypvarsCharged}
	\begin{aligned}
		\delta z_i{}^I &= 2\bar{\epsilon}_{ip}\zeta^{Ip}\,,\\
		\delta \zeta^{Ip} &= \slashed{\cD}z_i{}^I \epsilon^{ip} - \f12 C z_i{}^I \epsilon^{ip} + \f12 z_i{}^I\eta^{ip} + q_I L^p{}_q z_i{}^I \epsilon^{iq}\,,\\
		\delta \wti{z}_p^{\wti I} &= 2\bar{\epsilon}_{ip}\wti{\zeta}^{\wti I i}\,,\\
		\delta \wti{\zeta}^{\wti I i} &= \slashed{\cD}\wti{z}_p{}^{\wti I} \epsilon^{ip} + \f12 C \wti{z}_p{}^{\wti I} \epsilon^{ip} + \f12 \wti{z}_p^{\wti I}\eta^{ip} + \wti q_{\wti I} \wti L^i{}_j \wti z_p{}^{\wti I} \epsilon^{jp}\,.
	\end{aligned}
\end{equation}
%	

%%%%%%%%%%%%%%%%%%%%%%%%%%%%%%%%%%%%%
\section{1d sectors from the $\psu(2|2)\times\uu(1)$-invariant squashed sphere}
\label{app:anotherOne}
%%%%%%%%%%%%%%%%%%%%%%%%%%%%%%%%%%%%%

In Sections~\ref{sec:1dSectorCohomology} and~\ref{sec:1dSector}, we extensively discussed the one-dimensional sectors arising from $\cN=4$ three-dimensional QFTs in the $\su(2|1)\times \psu(1|1)$-invariant squashed sphere background. In Section~\ref{sec:background}, we introduced two inequivalent $\cN=4$ backgrounds, and therefore it is natural to ask whether the $\psu(2|2)\times \uu(1)$-invariant background preserves a similar 1d protected sector.

The $\psu(2|2)\times \uu(1)$-invariant background was introduced in Section~\ref{BACKGROUND2}, where we carefully analyzed its preserved supersymmetry algebra. With these results at hand we can proceed similarly as in the main text and look for 1d protected sectors in this background. As discussed there, the existence of such a 1d sector crucially relies on the existence of an $\su(1|1)$ subalgebra of the full supersymmetry algebra which is not contained in any $\cN=2$ subalgebra. To show that such an algebra exists, it is useful to introduce the following nilpotent supercharges, written in terms of the supercharges defined in \eqref{susies2}:
\begin{align}
	\cQ^H_1 &= Q_1^1 + \f{b^2}{1-b^2} Q_2^2 \,, & \cQ^H_2 &= \wti{Q}_1^1 + \f{1}{1-b^2} \wti{Q}_2^2\,,\\
	\cQ^H_3 &= Q_2^1 - \f{1}{1-b^2} \wti Q_2^1\,, & \cQ^H_4 &= Q_1^2 - \f{b^2}{1-b^2} \wti{Q}_1^2\,.
\end{align}
In this case we find two sets of two supercharges, $\cQ^H_1$ and $\cQ^H_2$ and $\cQ_H^3$ and $\cQ^H_3$ which both generate a $\psu(1|1)$ subalgebra. Following the strategy of Section \ref{sec:1dSectorCohomology}, we define the sums $\cQ^H = \cQ^H_1 + \cQ^H_2$ and $\widehat\cQ^H = \cQ^H_3 + \cQ^H_4$, which square to
\begin{equation}\label{anotherQQ}
	\begin{aligned}
		\acomm{\cQ^H}{\cQ^H} &= \f{8}{r}\left( i\,\bP_\beta + i\,\bH - \f{1+b^2}{1-b^2} Z_3 \right)\,,\\ \acomm{\widehat\cQ^H}{\widehat\cQ^H} &= \f{8}{r}\left( i\,\bP_\beta - i\,\bH - \f{1+b^2}{1-b^2} Z_3 \right)\,,
	\end{aligned}
\end{equation}
where, as before, $\bH$ is the Cartan generator of $\su(2)_H$, and $\beta=\f12(\psi-\phi)$ is defined as before. A local operator belongs to the cohomology of these supercharges if and only if the right-hand side of \eqref{anotherQQ} vanishes on this operator. Hence, we can again conclude that the operators in the cohomology have to be invariant under $\bH$ and $\bQ$ and have to be inserted along the circle at $\theta=0$ parameterized by $\alpha=\f12(\psi+\phi)$. As the operators of interest are invariant under $\bH$, the twisted translation reduces to a standard translation along the $\alpha$, i.e.~$i\, \widehat{\bP}_H= i \,\bP_\alpha\pm i \bH \simeq i \,\bP_\alpha$, where the last equality is understood to be valid only on the operators in the cohomology. Similar to the previous case we note that
\begin{align}
	\f{r}{8}\acomm{\cQ^H}{b^2 Q_1^1+\f{1}{1-b^2}Q_2^2} &= i\, \bP_\alpha + i\, \bH- \f{1+b^2}{1-b^2} Z_3 \,, \\
	\f{r}{8}\acomm{\widehat\cQ^H}{b^2 Q_1^2+\f{1}{b^2}Q_2^1} &= i\, \bP_\alpha - i\, \bH - \f{1+b^2}{1-b^2} Z_3 \,.
\end{align}
As all the operators in the cohomology are invariant under $\bH$ and $Z_3$, this translation is trivial at the level of the cohomology. In particular, the correlation functions cannot depend on the separation along the circle and the 1d theory is necessarily topological. To completely characterize the operators in the cohomology it therefore suffices to consider them inserted at $\alpha=0$.

From the cohomological construction it is clear that the twisted translated (local) operators in this case are much more constrained. As the hypermultiplet scalars are charged under $\bH$, the first place to look for local operators in the cohomology of $\cQ^H$ or $\widehat{\cQ}^H$ are the twisted hypermultiplets. Indeed, the twisted hypermultiplet scalars are uncharged under $\bH$ and $Z_3$. Again, for simplicity we will only consider a single twisted hypermultiplet but the generalization to more twisted hypermultiplets is clear. From the supersymmetry variations for the twisted hypermultiplets, it is easy to see that the linear combination
\begin{equation}
	\cW^I(\alpha) = \f{1}{\sqrt{1+b^2}}\left( b \,\wti z_1^I(\alpha) + \wti z_2^I(\alpha) \right)\,,
\end{equation}
is $\cQ^H$-invariant, and similarly
\begin{equation}
	\widehat{\cW}^I(\alpha) = \f{1}{\sqrt{b^2+b^{-2}-1}}\left( (b-b^{-1}) \wti z_1^I(\alpha) + \wti z_2^I(\alpha) \right)
\end{equation}
is invariant under $\widehat{\cQ}^H$. However, when computing the correlation functions, it becomes clear that all correlation functions between such operators are trivial in the cohomology, meaning that both $\cW$ and $\widehat{\cW}$ are not only $\cQ^H$- and $\widehat\cQ^H$-closed but also $\cQ^H$- and $\widehat\cQ^H$-exact.

More generally, one can exclude the appearance of any local operator with non-trivial correlations in this sector. Therefore we are led to conclude that in this case there is no interesting 1d sector of local operators. This is, however, not necessarily the end of the story, as there might be interesting non-local observables, such as Wilson lines or vortex lines, that are non-trivial in the cohomology. It would be very interesting to further investigate this case and see if such a protected sector exists.

%%%%%%%%%%%%%%%%%%	
\bibliography{AllBibs}
\bibliographystyle{ssg}
%%%%%%%%%%%%%%%%%%
	
\end{document}